\documentclass[preprint,12pt,authoryear]{elsarticle}

\usepackage{amssymb}
\usepackage{amsmath}
\DeclareMathOperator*{\minimize}{minimize}
\usepackage{amsthm}
\usepackage{float}
\usepackage{array}
\usepackage{comment}
\usepackage[noend]{algpseudocode}
\usepackage{algorithm,algorithmicx}
\makeatletter
\newenvironment{breakablealgorithm}
  {
   \begin{center}
     \refstepcounter{algorithm}
     \hrule height.8pt depth0pt \kern2pt
     \renewcommand{\caption}[2][\relax]{
       {\raggedright\textbf{\ALG@name~\thealgorithm} ##2\par}%
       \ifx\relax##1\relax 
         \addcontentsline{loa}{algorithm}{\protect\numberline{\thealgorithm}##2}%
       \else 
         \addcontentsline{loa}{algorithm}{\protect\numberline{\thealgorithm}##1}%
       \fi
       \kern2pt\hrule\kern2pt
     }
  }{
     \kern2pt\hrule\relax
   \end{center}
  }
\makeatother

\usepackage{multirow}
\usepackage{tabularx}
\usepackage{booktabs}
\usepackage{makecell}
\usepackage[margin=2cm]{geometry}
\usepackage{lineno,hyperref}
\usepackage{enumitem}
\usepackage{subfigure}
\usepackage{threeparttable}
\modulolinenumbers[5]
\hypersetup{
    colorlinks=true,
    linkcolor=blue,
    filecolor=blue,      
    urlcolor=blue,
    citecolor=cyan,
}

\usepackage{soul}  
\sethlcolor{yellow}

\usepackage{threeparttable}

\begin{document}

\begin{frontmatter}



\title{{Novel operational algorithms for} ride-pooling as on-demand feeder services}



\author[1]{Wenbo Fan \corref{cor1}} 
\ead{wbfan@swjtu.edu.cn}

\author[1]{Xiaotian Yan}
\ead{yanxiaotian@my.swjtu.edu.cn}

\author[1]{Zhanbo Sun \corref{cor1}}
\ead{zhanbo.sun@home.swjtu.edu.cn}

\author[1]{Xiaohui Yang}
\ead{yxh0709@my.swjtu.edu.cn}

\address[1]{Department of Traffic Engineering, Southwest Jiaotong University, Chengdu, China}

\cortext[cor1]{Corresponding author}

\begin{abstract}
 Ride-pooling (RP) service, as a form of shared mobility, enables multiple riders with similar itineraries to share the same vehicle and split the fee. This makes RP a promising on-demand feeder service for patrons with a common trip end in urban transportation. 
 We propose the RP as Feeder (RPaF) services with tailored operational algorithms. 
 Specifically, we have developed (\romannumeral 1) a batch-based matching algorithm that pools a batch of requests within an optimized buffer distance to each RP vehicle; (\romannumeral 2) a dispatching algorithm that adaptively dispatches vehicles to pick up the matched requests for certain occupancy target; and (\romannumeral 3) a repositioning algorithm that relocates vehicles to unmatched requests based on their level of urgency. We also embed the Traveling-Salesman-Problem (TSP) model to generate routing plans for each dispatched vehicle.
 An agent-based microscopic simulation platform is designed to execute these operational algorithms (via the Operator module), generate spatially distributed random requests (Patron module), and account for traffic conditions (Vehicle module) in street networks.
 Extensive numerical experiments are conducted to showcase the effectiveness of RPaF services across various demand scenarios {in typical morning rush hours}. 
 We compare RFaF with two on-demand feeder counterparts proposed in previous studies: Ride-Sharing as Feeder (RSaF) and Flexible-Route Feeder-Bus Transit (Flex-FBT). Comparisons reveal that {given the same fleet size, RPaF generally outperforms RSaF in higher service rates (i.e., the percentage of requests served over all requests) and Flex-FBT in shorter average trip times for patrons.} Lastly, we illustrate the implementation of RPaF in a real-world case study of the uptown Manhattan network (USA) using actual taxi trip data. {The results demonstrate that RPaF effectively balances the level of service (service rate and patrons' average trip time) with operational costs (fleet size).} {The proposed simulation platform offers a valuable testbed for evaluating and optimizing various on-demand feeder services.}
\end{abstract}




\end{frontmatter}


\section{Introduction}\label{sec:intro}
The rise in ride-hailing mobility has posed significant challenges to traditional urban public transportation systems, causing a global decline in ridership since the 1990s \citep{graehler2019, olayode2023}. The onset of the COVID-19 pandemic has exacerbated this trend, prompting a further shift from public transportation to individual mobility \citep{luo2022,shaheen2022}. Evidence shows that transit ridership in most cities worldwide has yet to recover to pre-pandemic levels \citep{ziedan2023}. This global phenomenon threatens the financial viability of many transit agencies, placing an increased burden on government subsidies.

To reverse this downward trend, some transit agencies are exploring innovative solutions to incorporate ride-hailing technologies into On-Demand Transit (ODT) services in hopes of luring back riders \citep{ghimire2024}. One of the major applications of these technology-enabled ODTs is to connect transportation hubs and their service zone(s) where demand is too scattered and low to support the operation of fixed bus lines \citep{koffman2004}. While ODT, sometimes referred to as demand-responsive transit or paratransit, is not a novel concept \citep{vuchic2007}, the integration of ride-hailing technologies -- such as smartphone-based wireless communication and global position systems (GPS) -- elevates the user experience through dynamic responses, location-based services, seamless payment, and transparent information. 

On the other hand, Transportation Network Companies (TNCs), as the platform provider of ride-hailing mobility, are introducing Ride-Pooling (RP) services as a first/last-mile solution to public transportation, either in partnership with transit agencies or independently \citep{shaheen2019}. RP services, such as Uber (Express) Pool, Lyft Line/Shuttle, Via Vanpooling, and Didi Express Pool, match multiple requests with similar itineraries to share the ride in the same vehicle. The RP not only improves transit accessibility as a faster feeder mode but also addresses the drawbacks of the conventional non-shared ride-hailing vehicles, which have been criticized for inducing individual trips, worsening traffic congestion, and increasing air pollution \citep{Alonso-Mora2017, FURUHATA201328, chan2012}.  

Both the technology-enabled ODTs and RP are On-Demand Shared-Ride (ODSR) mobility that offers customized door-to-door routing services. However, they differ in several operational aspects, as noted by \cite{fan2024optimal}: 
\begin{itemize}
    \item ODTs generally adhere to schedule/frequency-based operation strategies, either fixed for reserved requests or flexible for dynamic requests \citep{shahin2024survey}. Accordingly, ODTs dispatch vehicles \textit{sequentially} at headways and match requests within each headway interval to successive vehicles. In such a ``sequenced operation'', each ODT vehicle has to traverse the entire service zone where the matched requests are dispersed. 
    \item In contrast, RP vehicles are \textit{distributed} over the network and are matched with the {closest} requests, with dispatches triggered by successful matches. As a result, this ``distributed operation'' confines each RP vehicle's pickup process within the neighborhood area of its current location.
\end{itemize}


Therefore, we see RP's potential to lower pickup costs for operators and patrons. This motivates us to investigate: (\romannumeral 1) whether RP could surpass ODTs in overall system performance; (\romannumeral 2) under what conditions this might occur; and (\romannumeral 3) to what extent the advantage would manifest. 
To address these questions, we focus on the ``feeder problem'' (also called first-mile and last-mile problem) framework \citep[e.g.,][]{clarens1975, chang1991}, where feeder services are operated to transport patrons with dispersed locations to/from a common hub. The feeder-problem scenario applies to various contexts, where the hub can be a subway station, a school, a shopping mall, a stadium, a convention center, etc., and the travel demand is derived from commuters, students, shoppers, and audiences.\footnote{In principle, the framework also applies to logistics transportation such as last-mile delivery problems.} Several pioneering works have been done on designing ODTs in the feeder-problem framework \citep[e.g.,][]{CHEN201738, Luo2019b, sangveraphunsiri2022}. For comparison, we propose innovative operational algorithms for RP-as-feeder (RPaF) services. 

The main contributions of this paper are threefold:


\begin{enumerate}[label=(\roman*)]
    \item Three operational algorithms are tailored for RPaF services. They are (1) a batch-based matching algorithm that functions based on the buffer distance concept, ensuring that each RPaF vehicle is assigned requests within its neighborhood; (2) a dispatching algorithm that employs hard- and soft-target models to deploy vehicles according to fixed and variable occupancy targets, respectively; and (3) a repositioning algorithm that relocates vehicles to unmatched requests based on their level of urgency, which is quantified as a weighted sum of the time elapsed since the request was made and the distance between the request and the vehicle. 
    \item A microscopic simulation environment is developed to implement the proposed system as well as its ODSR counterparts. The simulation framework consists of three modules designed to capture the interactions among the operator, patrons, and vehicles. Such a module-based simulation framework serves as a testbed for evaluating the efficacy of the operational algorithms under diverse demand patterns in both stylized and real-world street networks.
    \item Systematic comparisons are conducted between the proposed RPaF and two counterparts, namely Ride-Sharing as Feeder (RSaF) \citep{DAGANZO2019213, LIU202122} and Flexible-Route Feeder-Bus Transit (Flex-FBT) \citep{chen2022, liu2020effects, EDWARDKIM201967, KIM2013227, Kim2012, chang1991}, which have been previously studied as on-demand feeder services. Their operational details are provided in Section \ref{counterparts}. Our study elucidates the impacts of their operational differences and delineates the application domains of RPaF through extensive experiments. Furthermore, a real-world case study contrasts the three ODSR modes with the current non-shared taxis.
\end{enumerate}

The remainder of this paper is organized as follows. The next section reviews related studies. Then, 
Section \ref{sec:Oper-algor} develops key operational algorithms for RPaF, which are embodied into the module-based simulation platform in Section \ref{sec:metho}.
Section \ref{sec:experi} presents numerical studies that examine the operational algorithms and compare alternative on-demand feeder systems. A real case study is demonstrated in Section \ref{sec:real} using taxi trip data in uptown Manhattan, New York, USA. Finally, Section \ref{sec:conclu} concludes this paper. Since there are several abbreviations, Table \ref{tab:Abbreviations} in \ref{appx_notation} lists them along with their meanings for the reader's convenience.

\section{Related studies}
Feeder services are essential in transit networks to enhance accessibility, facilitating passengers to travel between transit stations and their various origins and destinations. Research on feeder services is extensive and can be categorized by transportation modes \citep{malucelli1999demand, errico2013}. 
Traditional feeder modes comprise (\romannumeral 1) on-demand individual modes, e.g., walking, cycling (private or shared) bikes, driving (private or shared) cars, and non-shared ride-hailing taxis, and (\romannumeral 2) fixed shared-ride modes, e.g., Fixed Feeder-Bus Transit (Fix-FBT). However, these modes often exhibit notably poor performance in one or more of the three transportation aspects: mobility, accessibility, and efficiency, as summarized in Table \ref{tab:summary_feeder}. 

\begin{table}[!ht]
	\caption{\textbf{Qualitative comparisons between RPaF and other feeder modes}}
	\label{tab:summary_feeder}
        
	\vspace{0.1in}
	\centering
    \begin{threeparttable}
	\begin{tabularx}{\textwidth}{p{0.2\textwidth} X p{0.22\textwidth}}
		\hline
		\textbf{Feeder mode} & \textbf{Characteristics} &  \\
		\hline
        \makecell[l]{Ride-Pooling\\as Feeder (RPaF) } & \multicolumn{2}{l}{\makecell[l]{On-demand shared-ride door-to-door service; distributed operation\tnote{*};\\ vehicle occupancy of 2$\sim$7; private/dedicated drivers or autonomous driving; \\average in-vehicle speed of 15$\sim$50 km/h}}\\
    	\hline
        \textbf{Alternatives}\tnote{1} & \textbf{Characteristics} & \textbf{Versus RPaF}\\
    	\hline
    
		Walking & On-demand individual mobility; average speed of 2$\sim$5 km/h & Low speed (mobility) \\
	
		Cycling (private or shared) bikes & On-demand individual mobility; borrow and return shared bikes anytime, anywhere; average riding speed of 5$\sim$12 km/h & Riding eligibility required; low speed (mobility) \\
	
		Driving (private or shared) cars & On-demand individual mobility; one-way or two-way mobility; average riding speed of 15$\sim$60 km/h & Driving eligibility required; low vehicle occupancy (efficiency)\\
	
		Non-shared ride-hailing taxis & On-demand individual door-to-door service; distributed operation (matching; dispatching); dedicated drivers; average operation speed 15$\sim$60 km/h & Low vehicle occupancy (efficiency) \\
	
		Fixed feeder-bus & Traditional shared-ride stop-to-stop services; fixed routes, stops, schedules; vehicle occupancy of 2$\sim$25; average in-vehicle speed of 15$\sim$40 km/h &  Low accessibility \\

        On-Demand Transit (ODT) & On-demand shared-ride door-to-door service; flexible route/stop for door-to-door services; sequenced operation\tnote{**}; vehicle occupancy of 2$\sim$15; dedicated drivers or autonomous driving; average in-vehicle speed of 15$\sim$40 km/h & \textbf{To be studied in this paper}\\
        \hline
	\end{tabularx}
    
    \begin{tablenotes}
        \footnotesize
        \item[1] The literature on RP and ODT are discussed in detail as follows. For those of shared bikes, see \cite{luo2021, wu2020, li2023}. See \cite{aldaihani2004, CHEN201738, CHEN2017444, ramezani2023dynamic} for non-shared taxis. See \cite{jorge2015} for carsharing. And \cite{SIVAKUMARAN2014204} for fixed feeder-bus transit.
        \item[*] Distributed operation means that matching vehicles and requests based on their proximity in the network and successful matches determine the dispatches of the vehicles.
        \item[**] Sequenced operation means dispatching vehicles by headways and matching requests during the time interval to successive vehicles. 
    \end{tablenotes}
    \end{threeparttable}
\end{table}

Combining on-demand and shared-ride properties, ODT and RP aim to balance the above three inherently conflicting aspects. Particularly, RP/ODT-as-feeder services demonstrate potential by streamlining the matching and transportation of multiple requests with a common trip end, making them more efficient compared to serving many-to-many (M2M) demands \citep{errico2021}.
The state-of-the-art research of the two innovative feeder modes is summarized as follows. 
For general RP and ODT with M2M demands, interested readers can refer to recent surveys such as \cite{MOURAD2019323} and \cite{vansteenwegen2022}. 

We begin with the early development of ODSR systems that served as the precursor to modern ODT and RP.
    
\subsection{Early on-demand shared-ride feeder services}
The early ODSR system primarily consisted of Dial-A-Ride (DAR) services, which operated on a reservation basis depending on 20th-century communication technologies such as the telephone and the Internet. These services are commonly used by captive passengers with low travel frequency, including individuals with disabilities who require transportation to hospitals or workplaces.

DAR-as-feeder studies date back to \cite{wilson1976}, \cite{wilson1968}, and \cite{wilson1980}, and then developed by many scholars such as \cite{liaw1996}, \cite{hickman2001}, and \cite{aldaihani2003}. Due to its reservation-based low-frequency operational characteristics, DAR-as-feeder research primarily focuses on routing and scheduling problems for one-time out-of-depot vehicle trips. In this setting, each vehicle in the fleet is tasked with a single depot-based round trip for picking up or dropping off its assigned passengers within its capacity constraint. Constraints from the passengers' side are also respected such as their preferred time windows and maximum ride duration. The goal is usually to minimize the total vehicle distance traveled. These tactical operational problems are called static DAR Problems (DARP) with one-to-many (O2M) and many-to-one (M2O) demand. 
Variant models have been developed in the DARP framework, such as those employing different routing policies \citep{feng2014}, considering irregular street networks \citep{pan2015}, allowing meet points for multiple passengers boarding and alighting \citep{li2018}, and using alternative mathematical approaches \citep{yu2015}. Both exact and heuristic solution algorithms are developed for the proposed problems; {see \cite{cordeau2003dial} for an overview.} 

Dynamic DARPs have also been developed to tackle en-route instantaneous pickup requests or cancellations. The main trade-off is between the benefits/costs of the reserved passengers and last-minute requests. A common objective is to maximize the number of served passengers. Various approaches are proposed to facilitate the insertion of future requests into the existing schedule and routing plan. 
For instance, \cite{wang2020} modeled a two-step dynamic DARP, in which the first step constructs the initial routing and scheduling plan, and the second step modifies it to respond to real-time requests. 
{A thorough survey can be found in \cite{ho2018survey}.}

The static and dynamic DARP studies enhance fundamental theories for vehicle routing and scheduling of ODSR services. Yet, increasing demand driven by advanced communication technologies poses new operational challenges, necessitating ODSR vehicles to operate dynamically for multiple round trips between the hub and service areas. The management of recurrent vehicle trips determines whether ODSR evolves into ODT and RP under sequenced or distributed operations, respectively.

\subsection{ODT-as-feeder services}
ODT-as-feeder services can be further divided into two groups: (\romannumeral 1) flexible and (\romannumeral 2) semi-flexible feeder-bus transit, with the latter adhering partially to fixed routes/stops and timetables, permitting spontaneous deviations. 

\textit{(\romannumeral 1)  Flexible Feeder-Bus Transit (Flex-FBT):} Flex-FBT studies emerge in the 1970s with \cite{clarens1975} and \cite{ward1975}. To confine the vehicle distance traveled for serving dispersed demand, \textit{zoning} strategies have been utilized to divide the service area into smaller zones, each with dedicated buses. Flex-FBT research focuses on macroscopic system designs, rather than specific routing and scheduling plans, involving determinations of zone areas, headways, fleet sizes, and bus capacities. Analytical models have been developed to estimate system costs for operators and passengers. These models rely on approximations of average vehicle tour lengths for multiple pickups and drop-offs \citep{Stein1978, DAGANZO1978325, DAGANZO1984}.

The system-design modeling framework is adopted by subsequent studies. For instance, \cite{wirasinghe1977} jointly optimized a rail line and its Flex-FBT.
\cite{jacobson1980} compared the Flex-FBT system with direct door-to-door services and unveiled the crucial impact of transfer penalties on the performance of Flex-FBT.
\cite{adebisi1982} presented a comparative analysis for choosing between a Flex-FBT and a single fixed bus line according to the ridership levels.
\cite{chang1991a, chang1991} further compared Flex-FBT and Fix-FBT and accordingly proposed integrating the two options for lower- and higher-demand periods, respectively. 

The comparison between Flex- and Fix-FBT remains an attractive topic in subsequent studies such as \cite{edwards2013comparing} using real-world survey data, \cite{badia2021} using automated buses, and \cite{leffler2021simulation} in different network structures.
The analytical models obtained by \cite{chang1991a, chang1991} were then extended by several studies for varying contexts, such as mixed fleets \citep[e.g.,][]{Kim2012, KIM2013227}, coordinated schedules \citep[e.g.,][]{kim2014}, joint optimization of the service area and headway \citep[e.g.,][]{EDWARDKIM201967,chen2020solving} and auto buses \citep[e.g.,][]{guo2017, liu2020effects}. \cite{huang2019analytical} further employed the modeling framework to consider travelers' departure time choices. 

\textit{(\romannumeral 2) Semi-flexible Feeder-Bus Transit (Semiflex-FBT):} Semiflex-FBT, also called Mobility Allowance Shuttle Transit (MAST) or Demand Adaptive Transit (DAT), generally operates in a corridor following pre-specified routing policies. \cite{Welch1991} is among the earliest works to introduce the concept of Semiflex-FBT, highlighting the trade-off between increased accessibility of on-demand detours and the negative impact on the mobility of through riders.
\cite{fu2002} developed an analytical theory for Semiflex-FBT system designs regarding the optimal slack time between consecutive fixed stops to absorb random detours. A similar work is \cite{crainic2012}, who turned to design specific time windows at the fixed stops.
\cite{smith2003} further included the service zone size of Semiflex-FBT as the key design variable.

The work of \cite{quadrifoglio2006} has a significant impact on subsequent research. The author conducted an analytical analysis to approximate the Semiflex-FBT vehicle tour length under a specific routing strategy, known as the no-backtracking policy \citep{daganzo2005logistics}. The closed-form models obtained by \cite{quadrifoglio2006} laid the foundation for the optimal system design of Semiflex-FBT in several works \citep[e.g.,][]{zhao2008,quadrifoglio2009,li2011,chandra2013,papanikolaou2021,sangveraphunsiri2022}.
Alternative routing strategies have been explored, such as combinations of Fix/Flex-FBT, shortcuts, and short turns \citep{ceder2013integrated}, dynamic station strategy \citep{qiu2014,qiu2015exploration,qiu2015methodology,qiu2015demi}, route and point deviation strategies \citep{zheng2018methodology}, meeting point strategy \citep{zheng2019benefits}, and fixed route with flexible stopping schedule \citep{mehran2020}. 

It is worth noting that \cite{nourbakhsh2012} expanded Semiflex-FBT to a structured transit network for monocentric cities. Analytical models were developed for optimizing system designs concerning the service zone size and headway of each Semiflex-FBT line as well as a network structure parameter (i.e., the central zone area).
\cite{CHEN201738, chen2018} further integrated Semiflex-FBT with the fixed-route transit network and jointly optimized the integrated system in grid and radial street network structures.
\cite{calabro2023} advanced the integrated system design by relaxing the assumption of spatially uniform demand distribution in \cite{nourbakhsh2012} and \cite{CHEN201738, chen2018}.
In such integrated systems, \cite{leffler2024} introduced mode- and route-choice models for passengers to capture the day-to-day learning process and the demand splits among multiple transit modes.

In contrast to the above system design studies, attention was also paid to the vehicle routing problem of Semiflex-FBT as a variant of DARPs with partial fixed stops. \cite{crainic2005meta} tested several meta-heuristic solution algorithms (including the tabu search method and greedy multitrial randomized method) to solve the routing problem of a single-line Semiflex-FBT, which was later attacked using customized heuristic algorithms \citep{errico2021}. Other meta-heuristic algorithms have also been investigated, such as a multi-stage solution algorithm based on space partitioning and the genetic algorithm \citep{lu2016}, and a large neighborhood search algorithm \citep{montenegro2021}. A comprehensive survey is available in \cite{shahin2024survey}.


\subsection{RP-as-feeder services}
Technology-enabled RP contrasts with previous ODSR services primarily in its dynamic operational characteristics \citep{shen2018}. The ordering, matching, and dispatching of RP vehicles are executed in a distributed manner, without a specific sequence, and are completed in minutes or seconds. This dynamic and distributed context gives rise to new research topics, including the vehicle-request matching problem \citep{PaoloSanti2014, yang2020, ramezani2023dynamic}, the idling vehicle rebalancing problem \citep{Alonso-Mora2017, SIMONETTO2019208,valadkhani2023dynamic}, and fare splitting problem \citep{ke2020pricing, zhang2021pool}, besides the classic vehicle routing problem (formulated as dynamic DARPs) \citep{ho2018survey, namdarpour2024}.

RPaF is a variant of general-purpose RP for M2M demand \citep{AGATZ2012295, FURUHATA201328}. RPaF can be further divided into three groups according to who drives the vehicles: (\romannumeral 1) shared taxis with dedicated drivers hired by taxi companies or transit agencies, (\romannumeral 2) Peer-to-Peer (P2P) ride-sourcing/sharing, where private car owners participate in the crowd-sourced mobility platforms such as those of TNCs, and (\romannumeral 3) Autonomous Vehicles (AVs) that operate without human drivers. 

\textit{(\romannumeral 1) Shared Taxi-as-Feeder (STaF):} \cite{tao2007} is among the first to apply the shared-taxi concept \citep[previously developed in ][]{frattasi2005} to airport access. The STaF services are then evaluated by \cite{tao2008behavioral} and \cite{MA2017301} using empirical data. \cite{ma2019} considered STaF at both access and egress ends of transit trips and simulated for a case study of Long Island rail transit commuters to New York City.
\cite{bian2019a, bian2019b} conducted mechanism designs (including vehicle-request matching, vehicle routing, and pricing strategies) for STaF accounting for personalized requirements, such as extra in-vehicle travel time, the number of co-riders, and early-arrival waiting time at the transit hub.
\cite{zhu2020analysis} modeled shared taxis as both a feeder and an alternative to transit. They found the STaF-transit structure can benefit all three parties, i.e., the taxi operator, transit agency, and passengers.
{\cite{LIU202122} employed an overarching analytical framework for the joint design of STaF and transit systems. The design variables include the fleet size and zoning of STaF as well as the line spacing and headway of the transit network. They found that the integrated system can outperform the conventional transit systems fed by walking, Fix-FBT, and non-shared taxis.}

\textit{(\romannumeral 2) P2P Ride-Sharing-as-Feeder (RSaF):} The vehicle routing problems of P2P-RSaF are a novel variant of DARP, which involves private drivers' heterogeneous travel preferences and were initially studied by \cite{masoud2017,stiglic2018}. 
{\cite{kumar2021} suggested integrating P2P-RSaF with a schedule-based transit system. A rolling horizon approach was applied to solve dynamic ride-sharing with transfers at transit stops.}
\cite{nam2018} resorted to simulation tools to examine the effects of P2P-RSaF on the Los Angeles Metro Red Line subway rail. However, P2P-RSaF may not be a viable feeder mode for the general public because private drivers may impose stringent constraints and not comply with scheduling requirements. These issues can be addressed by leveraging AVs. 

\textit{(\romannumeral 3) Shared AVs-as-Feeder (SAVaF):} 
Driven by advancements in AV technology, the role of SAVaF in transit systems has been recognized \citep{shen2018,wen2018}.
For instance, \cite{shen2018} simulated SAVaF to replace low-demand Fix-FBT routes for the subway in Singapore. \cite{wen2018} simulated SAVaF among other feeder modes to rail stations. 
\cite{chen2020solving} studied SAVaF to a single subway station and formulated a static DARP for each epoch in a rolling horizon framework. 
\cite{Gurumurthy2020} and \cite{huang2021} simulated and compared integrated transit systems fed by SAVaF and direct door-to-door services in Austin (Texas, USA).
\cite{Pinto2020} simulated a transit network with SAVaF and jointly optimized the transit service frequency and the fleet size of AVs.
{\cite{imhof2020} evaluated the feasibility of SAVaF connecting railway stations in the rural area of the Toss Valley, Switzerland.}
{\cite{lau2021} estimated the transit demand brought by SAVaF considering travelers’ mode choice preferences in Kuala Lumpur (Malaysia).}
\cite{whitmore2022} examined the cost efficiency of integrating SAVaF for equitable transit coverage in Allegheny County, Pennsylvania (USA).





\subsection{Research gap}
We now summarize the above literature review and then identify research gaps to be addressed in this paper.

Four key points can be drawn from the literature review:
\begin{itemize}
    \item Early research on ODSR services surrounds the development of DARP theories, resulting in fruitful outcomes such as variant context-specific DARP models, exact solution algorithms, and (meta-)heuristic solution algorithms. These achievements have established the groundwork for vehicle routing and scheduling for both ODT- and RP-as-feeder services.
    \item The advance in ride-hailing technologies induced dynamic demand and transformed ODSR into ODT- and RP-as-feeder, which inherit operational features from traditional transit and taxi services, respectively. The former adopts zoning strategies and employs cyclical headway-based operation for each zone, while the latter intercepts trip requests based on temporal-spatial proximity to distributed vehicles in the network.
    \item Research on ODT-as-feeder has focused on system designs using approximation-based analytical models. System performance evaluations have demonstrated that ODT-as-feeder services can offer substantial advantages over traditional feeder modes (e.g., walking, non-shared taxis, and Fix-FBT) for low-density demand situations.
    \item Much effort of RP-as-feeder research has been devoted to addressing the new operational challenges, such as dynamic vehicle-request matching and vehicle rebalancing among others. Existing studies on the performance evaluation of RP-as-feeder services mostly involve early-stage concept designs, scenario simulations, and analytical analyses for optimal system designs. RP-as-feeder also proves beneficial in low-density demand scenarios.
\end{itemize}

As seen, both ODT- and RP-as-feeder exhibit advantages over traditional feeder modes in low-density demand scenarios. It is natural to question which of the two performs better and the application ranges distinguishing them. However, the existing literature lacks systematic comparisons between the two feeder modes, overlooking the impact of their subtly distinct operational features. An exception is \cite{fan2024optimal}, which compared a novel RP variant, RP with the Hold-Dispatch strategy (RP-HD), and Flex-FBT. Analytical models were developed for optimizing system designs under non-uniform demand distributions. Extensive numerical experiments demonstrated that RP-as-feeder services consistently outperform Flex-FBT in all test scenarios except when the fleet cost/patrons’ value of time becomes much higher/lower.

{Nevertheless, \cite{fan2024optimal} is a macro-level study based on aspatial queueing models that overlooked network structures and operational details like vehicle-request matching, vehicle dispatching, routing, and repositioning. This study will address these aspects in specific street networks and devise sophisticated operational algorithms that were previously ignored in the cited study for tractability.} Building on the research by \cite{fan2024optimal}, we aim to delve deeper into the efficacy of RPaF in stochastic and dynamic circumstances. To achieve this, we employ a simulation-based approach. Simulation models serve as ideal tools to execute theoretical service designs in realistic situations accounting for the impacts of many random factors. Laboratory experiments can be easily conducted across various scenarios to offer guidance for real-world applications. 

Numerous simulation-based tools have been developed for the operation of ODSR-as-feeder systems, such as customized micro-simulation platforms for DAR \citep{wilson1980}, Flex-FBT \citep{ceder2013integrated,edwards2013comparing,leffler2021simulation}, Semiflex-FBT \citep{horn2002fleet,horn2002multi,yoon2022simulation,leffler2024}, STaF \citep{ma2019}, P2P-RSaF \citep{masoud2017,kumar2021}, and SAVaF \citep{wen2018, Pinto2020}, along with open-source multi-agent simulation software like MATSim \citep{Gurumurthy2020} and SUMO \citep{huang2021} for SAVaF.
This paper adds a new tool to the arsenal for simulating RPaF services with innovative operational algorithms described below.

\section{RPaF with novel operational algorithms}\label{sec:Oper-algor}
{We begin with the RPaF service description in Section \ref{subsec:con and assump}, followed by novel operational algorithms in Section \ref{sec_algorithms}.} {In line with existing on-demand feeder systems (e.g., RSaF and Flex-FBT), we focus on dynamic, and spontaneous (rather than reservation-based) trips, where patrons typically indicate only their desired pick-up and/or drop-off locations.}\footnote{{Patrons' individual preferences, such as desired pick-up and drop-off times, are not explicitly considered in this paper. This setting enables us to derive analytical results for optimally parameterizing the matching algorithm in Section \ref{subsubsec_matching}. Potential extensions are discussed in Section \ref{sec:conclu}.}} 
The notations used in this paper are summarized in Table \ref{tab:notation} of \ref{appx_notation}.

\subsection{RPaF service description}\label{subsec:con and assump}
\begin{figure}[!ht]
 		\centering
 		\includegraphics[width=0.9\textwidth]{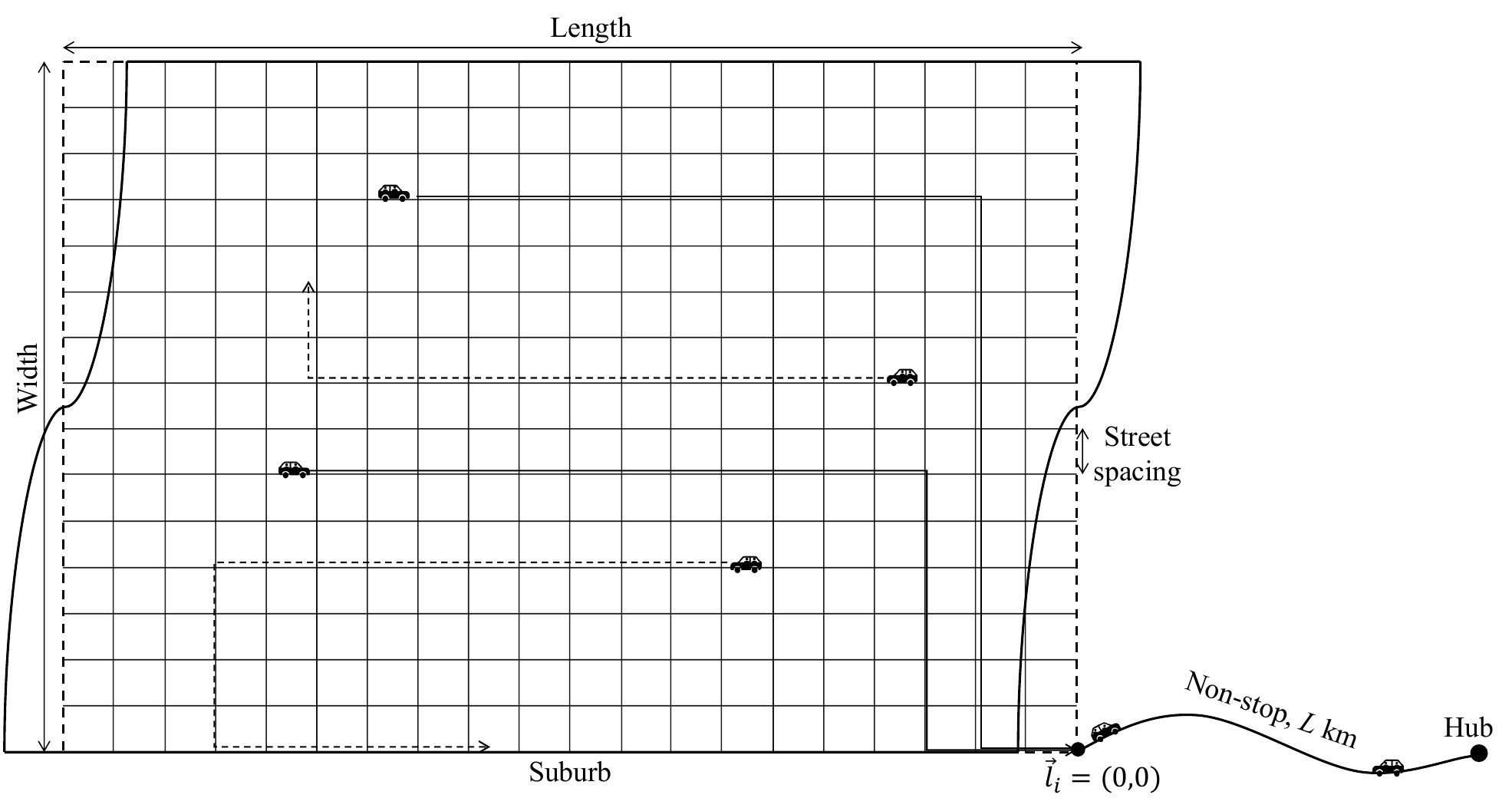}
		\caption{Illustration of ride-pooling as on-demand feeder services.}
		\label{fig:layout}
    \end{figure}

{We adopt the RPaF concept from \citep{fan2024optimal}, which connects a remote suburb and a downtown hub, as illustrated in Figure \ref{fig:layout}.} Similar layouts have been used in previous studies to study feeder transit systems, e.g., \cite{EDWARDKIM201967, KIM2013227}. {We use this single-hub system to ensure comparability with prior research. In cases of networks with multiple hubs, the proposed models can be utilized to operate separate RPaF services for each hub.} 


As shown in Figure \ref{fig:layout}, we first represent the suburb as a rectangular suburban region with an evenly spaced grid street network. (Later, we will apply the simulation framework to a case study using the real-world street network and demand data.)
We aim to examine a stabilized system, where the roadway network is configured with the following traffic characteristics. The cruising speed of RPaF vehicles is $S'$ [km/h] on suburban streets and $S$ [km/h] on freeways. {At intersections, RPaF vehicles experience delays dependent on control rules such as no control and signalized controls.} In addition, vehicles incur {a fixed delay $t_{d}$ [seconds]} for each stop to pick up or drop off. To enhance simulation efficiency, we disregard the dynamics of background traffic, which may have a minor impact in suburban areas. The simulation of dynamic traffic could be integrated into our framework as a potential future research direction. 

The travel demand between the suburb and the hub is bidirectional, denoted $\lambda_{\rm OB}$ [patrons/sq. km/hour] for outbound trips (from residences in the suburb to the hub) and $\lambda_{\rm IB}$ [patrons/sq. km/hour] inbound trips (from the hub to the suburb). We assume for now that travelers' origins and destinations are uniformly distributed throughout the suburban area. This assumption will later be lifted in the real case study.
Following previous studies (e.g., \citealp{DAGANZO2019213, EDWARDKIM201967, LIU202122}), we also assume that the demand is known and inelastic to RPaF service changes. (Induced demand can also be integrated into the simulation framework, but this is left for future study.) 

We will examine a single RPaF operator who oversees a fleet of e-hailing vehicles with {fully compliant drivers.} This operator could be a TNC, {a transit agency}, or a third-party integrator that combines services from various providers onto a unified platform. This setup reduces fierce competition among different operators and allows for pooling multiple requests without the concern of losing them to other competitors. {Therefore, we suppose travelers do not cancel their orders within a maximum tolerance time $\Bar{\tau}$ or after the vehicles are dispatched.} 


\subsection{Operational algorithms} \label{sec_algorithms}
This section develops four key operational algorithms for the above RPaF service, i.e., the matching, dispatching, routing, and repositioning algorithms.
\subsubsection{Matching algorithm}\label{subsubsec_matching}
The matching algorithm is deployed to match the spatially distributed (outbound) vehicles and patrons. Note that inbound vehicles and patrons meet at the hub's pick-up area and thus do not need online matching. Therefore, the subsequent discussion of the matching algorithm refers to outbound service.

Following \citep{MA2017301,yang2020, WEI2022100058}, we adapt the ``batch-based matching'' algorithm by using a target number of requests $U$ instead of a pre-specified time window, which {will be explained soon.} This algorithm pools $U$ requests into batches for each available vehicle within its matching buffer distance of $\Delta$ [km].
The algorithm then assigns the vehicles with the $u$ closest requests out of $U$. The unmatched requests (\(U-u\)) in the last batch are prioritized over new incoming requests in the following matching round. The steps of the batch-matching algorithm are outlined in Algorithm~\ref{algorithm_batch}. {The matching algorithm outputs a vehicle-request mapping dictionary $\mathbf{X}=\{X_n\}$.}

{The proposed algorithm differs from most existing ones, which depend on integer programming with bipartite graphs \citep[see, e.g.,][]{ramezani2023dynamic}. The distinction, however, enables our algorithm to yield the following valuable analytical properties absent in mathematical programming models and consequently reduces the computational complexity significantly.}

	
	\begin{algorithm}[!ht]
		\caption{{Batch-based matching algorithm}}
		\label{algorithm_batch}
		\hspace*{0.02in} {\bf Input:} 
		At the current simulation time $t$, request set $ R \equiv \left\{r_i = \langle t_i, \Vec{l}_i,\delta_i \rangle \right\} $, vehicle set $ V \equiv \left\{ v_n = \langle o_n, \Vec{l}_n, X_n, \delta_n \rangle \right\} $, matching buffer distance $\Delta$;\\
		\hspace*{0.02in} {\bf Output:} 
		At the current simulation time $t$, vehicle-demand mapping dictionary $\mathbf{X} \equiv \{X_n\}$.
		\begin{algorithmic}[1]
			\State Attain the last simulation time $ \mathbf{X}=\{X_n\} $.
			\For{each vehicle $n$ in $ V $} 
			\If{vehicle state $ v_n=\langle 0, \Vec{l}_n, X_n, \text{OB} \rangle $ and $|X_n| < u$ state}
			\State Let vehicle $n$ be the candidate vehicle.
			\State Calculate the remaining vehicle capacity $u-|X_n|$.
			\EndIf
			\State Define set $\mathbf{D}_n=\{d_{n,i}\}$ where $\{d_{n,i}\}$ denotes the distance between the \\ candidate vehicle $n$ and the candidate patron $i$ in $R$.
			\For{each patron $i$ in $ R $}
			\If{patron state $r_i=\langle t_i,\Vec{l}_i,0 \rangle$}
			\State Calculate distance $ d_{n,i} $ between vehicle $ n $ and patron $ i $.
            \hspace{0.85in}\makebox[0pt][r]{\hfill \raisebox{.5\baselineskip}[0pt][0pt]{$\left.\rule{0pt}{2.4\baselineskip}\right\}\ \mbox{in parallel}$}}

			\If{$ d_{n,i} \leq \Delta$}
            \hfill \raisebox{.5\baselineskip}[0pt][0pt]{$\left.\rule{0pt}{9.2\baselineskip}\right\}\ \mbox{in parallel}$}
			\State $\mathbf{D}_n \leftarrow \mathbf{D}_n \cup \{d_{n,i}\}$.
			\EndIf
			\EndIf
			\EndFor
			\State $m=\text {min} \left(u-|X_n|,\vert \mathbf{D}_n \vert \right)$. 
			\State $\mathbf{D}_{n}^{*}={\rm sort}(\mathbf{D}_{n}=\{d_{n,i}\})$ in increasing distance order.
			\State Filter the first $m$ elements($\{d_{n,i}^{*}\}$) from the $\mathbf{D}_{n}^{*}$.
			\State $X_n \leftarrow X_n \cup \{i\}$, where $\{i\}$ corresponds to $\{d_{n,i}^{*}\}$.
            \State $r_{i}=\langle t_i,\Vec{l}_i,1\rangle$, where $\{i\}$ corresponds to $\{d_{n,i}^{*}\}$.
            \State $v_{n}\leftarrow v_n=\langle 0, \Vec{l}_n, X_n, \text{OB}\rangle$.
			\EndFor
			\State \Return $ \mathbf{X}=\{X_n\} $.
		\end{algorithmic}
	\end{algorithm}	

The choice of parameters $U$ and $\Delta$ is crucial for the algorithm. {We propose the following recipe to determine proper values that minimize patrons' time delays in matching and subsequent routing processes.} First, we notice that the matching buffer distance $\Delta$ defines a coverage area of $A(\Delta)$ [km$^2$] for each available vehicle, e.g., $A(\Delta)=2\Delta^2$ in the Manhattan metric or $A(\Delta)=\pi \Delta^2$ in the Euclidean metric.\footnote{{The matching buffer distance $\Delta$ would be truncated to not exceed half the distance to the nearest request-receiving vehicle. Thus, vehicles' matching buffer areas would not overlap.}} In this area, pooling \(U\) requests takes an average time of 
\begin{align}
    \frac{U - \left(U-u\right)}{\lambda_{\rm OB} A(\Delta)}=\frac{u}{\lambda_{\rm OB} A(\Delta)},
\end{align}
where \(\left(U-u\right)\) is the number of requests left from the last batch and only \(u\) new requests need to be pooled per batch. 

For patrons in the current batch and those in the following, they experience respectively half and the entire pooling time for $U$ requests. Thus, their total pooling time delays can be estimated by
\begin{align} \label{apprpx_pooling_time}
    u\frac{u}{2\lambda_{\rm OB} A(\Delta)}+\left(U-u\right)\frac{u}{\lambda_{\rm OB} A(\Delta)}.
\end{align}

\begin{figure}[!ht]
		\centering
 		\includegraphics[width=0.55\textwidth]{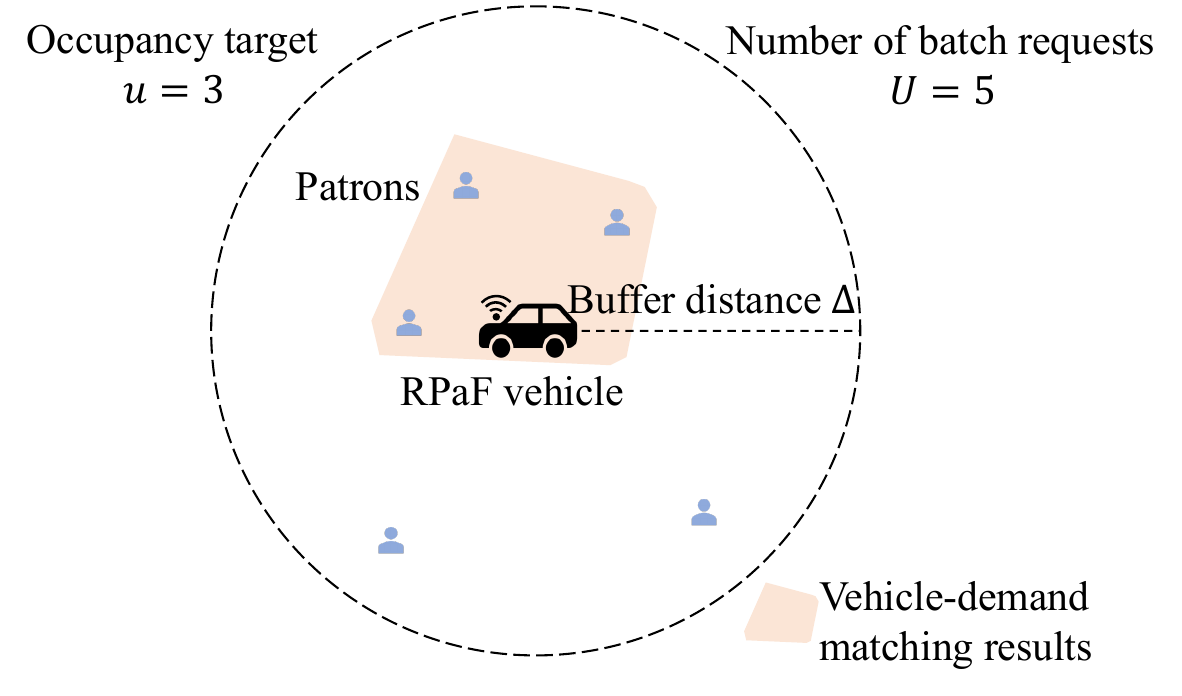}
		\caption{Illustration of ``batch-based matching''(filtering $u$ closest-to-the-vehicle requests out of $U$). 
		}\label{fig:matching_buffer}
	\end{figure}

In addition, the average travel time for visiting \(u\) points within an area of $A(\Delta)$ can be approximated by
\begin{align} \label{approx_trip_time}
    \frac{k\sqrt{A(\Delta)u}}{S'(u)}\frac{u}{U+1},
\end{align}
where the expected tour distance of traveling-salesman-problem (TSP) with $u$ visiting points, i.e., $k\sqrt{A(\Delta)u}$ according to \cite{Stein1978}, is reduced by a factor of $\frac{u}{U+1}$, capturing the effect of filtering $u$ closest-to-the-vehicle requests out of $U$, as illustrated in Figure \ref{fig:matching_buffer}.\footnote{A proof is given in \ref{appdx_u-U}.} The $k$ is a constant parameter, e.g., \(k=1.15\) in Manhattan metric and \(k=0.90\) in in Euclidean metric. The $S'(u)$ is the vehicles' commercial speed and satisfies $\frac{1}{S'(u)} = \left(\frac{1}{S'} + u t_{d}\right)$ considering stopping delays $t_{d}$.

Thus, patrons' total TSP-tour delays, encompassing both out-of-vehicle and in-vehicle waiting times, can be estimated by
\begin{align} \label{approx_TSP_time}
    u \frac{k\sqrt{A(\Delta)u}}{S'(u)}\frac{u}{U+1} + (U-u) \frac{k\sqrt{A(\Delta)u}}{S'(u)}\frac{u}{U+1} = U\frac{k\sqrt{A(\Delta)u}}{S'(u)}\frac{u}{U+1},
\end{align}
where the first term accounts for the first-matched $u$ patrons' TSP-tour delays, and the second for the next-matched $(U-u)$ ones', {which occur in their respective and independent TSP tours}. {Note that the next matching occurs with another nearby vehicle and is independent of the completion of the first-matched vehicle's pick-up task. We also note that the second term in Eq. (\ref{approx_TSP_time}), i.e., $(U-u) \frac{k\sqrt{A(\Delta)u}}{S'(u)}\frac{u}{U+1}$, is an approximation. It assumes that the reduction effect of ``filtering $u$ closest-to-the-vehicle requests out of $U$'' is not compromised by the prioritized $(U-u)$ requests (which may not belong to the $u$ closest-to-the-vehicle requests) in the next matching pool. Nonetheless, this approximation term will cancel out when $U=u$, making Eq. (4) accurate, which will be met by the following optimization.}

Adding Eqs. (\ref{apprpx_pooling_time}) and (\ref{approx_TSP_time})  yields total delays in the pooling and picking-up processes that are directly dependent on $U$ and $\Delta$:
\begin{align} \label{approx_total_time}
    U\frac{k\sqrt{A(\Delta)u}}{S'(u)}\frac{u}{U+1} + u\frac{u}{2\lambda_{\rm OB} A(\Delta)}+\left(U-u\right)\frac{u}{\lambda_{\rm OB} A(\Delta)}.
\end{align}

We now determine $U$ and $\Delta$ by minimizing Eq. (\ref{approx_total_time}): 
\begin{subequations} \label{appex_optimal_prob}
    \begin{align}
        \minimize_{U,A(\Delta)} U\frac{k\sqrt{A(\Delta)u}}{S'(u)}\frac{u}{U+1}-\frac{u^2}{2\lambda_{\rm OB} A(\Delta)}+\frac{uU}{\lambda_{\rm OB} A(\Delta)}
    \end{align}
    subject to:
    \begin{align}
        U \ge u, A(\Delta)>0.
    \end{align}
\end{subequations}
Through a few steps derivation according to the first-order condition, problem (\ref{appex_optimal_prob}) yields the following closed-form solutions:
\begin{align} \label{optimal_U}
        U^* = u,A^{*}(\Delta)=u^{-\frac{1}{3}}\left[ \frac{\left(u+1\right)S^{'}(u)}{k\lambda_{\rm OB}}\right]^{\frac{2}{3}},
\end{align}
implying the optimal matching buffer distance to be
\begin{align} \label{optimal_Delta}
    {\Delta^*} = 
    \begin{cases}
        & \left(8u\right)^{-\frac{1}{6}}\left[ \frac{\left(u+1\right)S^{'}(u)}{1.15\lambda_{\rm OB}} \right]^{\frac{1}{3}} \approx 0.675u^{-\frac{1}{6}}\left(\frac{\left(u+1\right)S'(u)}{\lambda_{\rm OB}}\right)^{\frac{1}{3}}, \text{ in Manhattan metric} \\
        & \left(\pi^{3} u\right)^{-\frac{1}{6}}\left[ \frac{\left(u+1\right)S^{'}(u)}{0.9\lambda_{\rm OB}} \right]^{\frac{1}{3}} \approx 0.584u^{-\frac{1}{6}}\left(\frac{\left(u+1\right)S'(u)}{\lambda_{\rm OB}}\right)^{\frac{1}{3}}, \text{ in Euclidean metric}.
    \end{cases}
\end{align}

\textbf{Remark}: The above results indicate that the optimal batch size should be identical to the occupancy target $u$. In other words, no one should be left unmatched to the next batch. {This result also explains why a time-window-based batch matching would probably be less effective, as $U \ge u$ would be generated in one or multiple pooling time windows until vehicles are dispatched upon at least $u$ matched requests.} Another finding from Eq. (\ref{optimal_Delta}) is that the optimal matching buffer distance $\Delta^*$ should increase/decrease as the demand density falls/rises with a power of $-\frac{1}{3}$.\footnote{{The optimization of a matching distance threshold is also discussed in \cite{ramezani2023dynamic}, which presents a closed-form approximation derived from regression analysis of simulated data. However, the reference primarily addresses non-shared ride-sourcing services.}}  Eq. (\ref{optimal_Delta}) also implies that larger $u$'s require longer matching buffer distances to shorten the pooling time, which outweighs the increase in TSP-tour time. 

{For those unmatched requests (with $\delta_i = 0 $), e.g., those out of the coverage areas of any vehicles, we sort them in the set \(\widetilde{X}_0\) (of a virtual vehicle indexed by $0$) in an increasing order of their call-in times. They will be visited by the vehicles designated by the repositioning algorithm. It is noted that the relocated vehicles may encounter $U>u$ awaiting requests in their matching buffer distances upon their arrival (i.e., the beginning of matching). In this over-saturated case, the matching to the relocated vehicles will be performed based on requests' ``urgency'', which will be soon defined in the repositioning algorithm. The \(\widetilde{X}_0\) is updated upon each match by removing $\{i|x_{n, i}=1, \forall n>0\}$.}

\subsubsection{Dispatching algorithm}\label{subsubsec:dispatching}
{This algorithm also functions for the outbound service since inbound vehicles follow the stop-and-go rule, as described in subsection \ref{sec_vehicle_module}.}
Following the idea of \cite{fan2024optimal}, we first incorporate a ``hard-target'' dispatch strategy that dispatches vehicles when the number of request assignments reaches the pre-given occupancy target, i.e., \(|X_n| = u \). The steps are outlined in Algorithm \ref{algorithm:disp}. 
    \begin{algorithm}[!ht]
        \caption{{Dispatching algorithm}}
        \label{algorithm:disp}
        \hspace*{0.02in} {\bf Input:}
        At the current simulation time $t$, vehicles set $V \equiv \left\{ v_n=\langle o_n, \Vec{l}_n, X_n, \delta_n \rangle \right\}$, vehicle-demand mapping dictionary $\mathbf{X} \equiv \{X_{n}\}$.
        \begin{algorithmic}[1]
            \State Attain $V_n$ and $\mathbf{X}$.
                \For{each vehicle $n$ in $V$}
                    \State Attain $X_n \in \mathbf{X}$.
                    \If{$ X_n=\emptyset $ state}
                        \State The vehicle $n$ heads to the assigned location (under the instructions of \\ the repositioning algorithm in Algorithm \ref{algorithm:reposition}).
                    \ElsIf{$|X_n|=1$}
                        \If{dispatch interval $t_n$ = {None}}
                            \State Set $t_n=t+\Bar{\tau}$
                            \hfill \raisebox{.5\baselineskip}[0pt][0pt]{$\left.\rule{0pt}{6.4\baselineskip}\right\}\ \mbox{in parallel}$}
                        \EndIf
                    \ElsIf{$ 1<|X_n|<u $ and departure interval $t_{n}$ is not reached}
                        \State The vehicle $n$ remains stationary.
                    \ElsIf{$|X_n|=u$ or $t$ reach departure interval $t_{n}$}
                        \State The vehicle $n$ is dispatched to the first request point followed by the \\ routing algorithm.
                    \EndIf
                \EndFor
        \end{algorithmic}
        
    \end{algorithm}
As found by \cite{fan2024}, the hard-target dispatch strategy outperforms its counterparts, e.g., a ``quick-dispatch'' strategy in \cite{DAGANZO2019213} and a headway-based strategy (of which the details in given in \ref{appdx-QD} and \ref{appdx-flex}) under medium to high levels of demand.

\begin{figure}[!ht]
    \centering
    \includegraphics[width=0.5\textwidth]{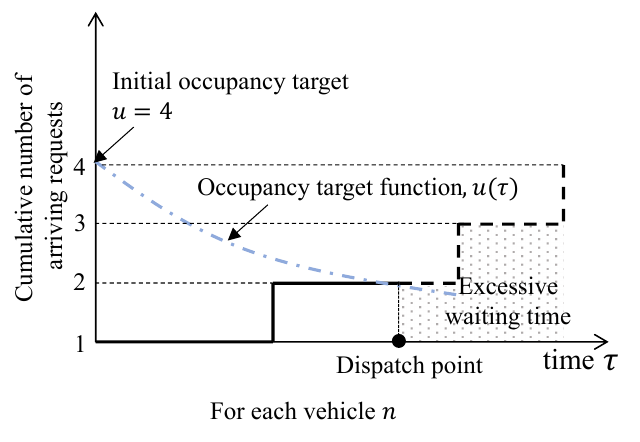}\label{fig:soft-RP}
	\caption{Illustration of soft-target dispatch strategy in RPaF.}\label{fig:soft-target}
\end{figure}

{Nonetheless, as noticed by \cite{daganzo2019public}, the hard-target dispatch strategy may cause prolonged wait delays to the arrived requests due to the late arrivals of the last few requests for a full match of $u$ requests. To address this issue, we modify the hard target to be a ``soft target'', with slightly abused notation \(u(\tau)\), that decreases gradually as the time \(\tau\) elapses since the vehicle's first request assignment, as illustrated in Figure \ref{fig:soft-target}.}

Several forms of time-declining models have been proposed for \(u(\tau)\), of which a simple one is as follows \citep{fan2024}:
\begin{align} \label{eq_soft_target}
    u(\tau) = \min \left\{\frac{2c_f}{\beta \tau}, u\right\},
\end{align}
where $u$ is the original fixed hard target; $c_f$ is the operational cost per dispatch; and $\beta$ denotes patrons' value of time (VOT). The derivation of Eq. (\ref{eq_soft_target}) can be found in \cite{daganzo2019public, fan2024} and thus, is omitted here. It is worth noting that \cite{fan2024} found that the soft-target strategy performs better for shuttle transit systems under demand distributions with large coefficients of variation, and the hard-target strategy is suitable for patrons with low VOTs (or equivalently when operational cost is high). 
This paper will examine two dispatching strategies for RPaF services.

{In addition, we will modify the dispatching algorithms to include a maximum interval since the first request arrived in the matching pool. The modification aims to mitigate the request-canceling issue to prevent excessive cancellations that could harm customer satisfaction and reduce demand. This maximum dispatch interval is set equal to patrons' maximum tolerance time $\Bar{\tau}$.}

\subsubsection{Routing algorithm}\label{subsubsec:routing}
Upon a dispatch order to vehicle \(n\), the routing algorithm is activated with the input of the match set \(X_n\). An optimization is performed to solve the optimal tour for visiting $|X_n|$ points plus the starting point from the vehicle's current location. The corresponding problem belongs to the classic open-tour TSP, which is provided in \ref{open-tour tsp} for the sake of completeness. The output is the routing plan $\Upsilon_n$ containing the sequence of visiting points {$\{\Vec{l}_{i \in X_n}\}$.}

Note that the TSP is small-scale since $|X_n| \le C $, where vehicle capacity $C=4$ for sedans, and independent for each vehicle, and the exact solution can be efficiently and dynamically found by off-shelf commercial algorithms, e.g., dynamic programming.

After completing the planned route, vehicle $n$, if of outbound service, heads to the hub; otherwise, the vehicle of inbound service follows the order from the repositioning algorithm. 


\subsubsection{Repositioning algorithm}
{This algorithm checks the unmatched request set $\widetilde{X}_0 \neq \emptyset$ and relocates the vehicles that have just finished delivering inbound patrons to the most ``urgent'' unmatched outbound request. The ``urgency'' of requests is weighted by the unmatched time and distance to the available vehicle $n$, as given by}
\begin{align} \label{urgency}
    \omega_{n, i} = \alpha\times\text{unmatched time} - (1-\alpha)\times \frac{\text{distance to the available vehicle } n}{S'}, \forall i \in \widetilde{X}_0,
\end{align}
{where $\alpha \in [0,1]$ is a weight coefficient specified by the operator.}

{For those inbound vehicles with $\tilde{o}_n=0$, the repositioning algorithm directs them to the location of the top request in unmatched set $\widetilde{X}_0$. }

{If $\widetilde{X}_0 = \emptyset$, vehicles that have just finished delivering inbound patrons stay at the last drop-off point; and vehicles with $\tilde{o}_n=0$ at the hub return to the last pick-up location in the outbound service tour.}

The detailed steps are given in Algorithm \ref{algorithm:reposition}.
\begin{algorithm}[!ht]
        \caption{{Repositioning algorithm}}
        \label{algorithm:reposition}
        \hspace*{0.02in} {\bf Input:}
        At the current simulation time $t$, vehicles set $V \equiv \left\{ v_n=\langle o_n, \Vec{l}_n, X_n, \delta_n \rangle \right\}$, and unmatched requests set $\widetilde{X}_0$.
        \begin{algorithmic}[1]
            \State Attain $V$ and $\widetilde{X}_0$.
                \For{each vehicle $n$ in Set $V$}
                    \State Attain state $v_{n}=\langle \tilde{o}_n,\Vec{l}_n,X_n,\delta_n=\text{IB} \rangle$.
                    \If{$v_{n}=\langle \tilde{o}_n=0,\Vec{l}_n,X_n,\delta_n=\text{IB} \rangle$}
                        \If{$\widetilde{X}_0 \neq \emptyset$}
                            \State Vehicle $n$ is dispatched to the $\text{max}\left\{\omega_{n,i}\right\}$ patron position (the first \\ patron in the $\widetilde{X}_0$ set).
                        \ElsIf{$\widetilde{X}_0 = \emptyset$}
                            \If{vehicle $n$ is at the hub}
                            \hfill \raisebox{.5\baselineskip}[0pt][0pt]{$\left.\rule{0pt}{7.0\baselineskip}\right\}\ \mbox{in parallel}$}
                                \State Vehicle $n$ has not received inbound patrons and returns to the \\ boarding location of the last pick-up patron in the previous outbound service.
                            \ElsIf{vehicle $n$ is at the suburb}
                                \State Vehicle $n$ has just finished delivering inbound patrons and is  \\ waiting for follow-up instructions at the last drop-off point.
                            \EndIf
                            \State $v_n \leftarrow v_n=\langle o_n,\Vec{l}_n,X_n,\delta_n=\text{OB} \rangle$.
                        \EndIf
                    \EndIf
                \EndFor
        \end{algorithmic}
        
    \end{algorithm}

\section{Simulation framework}\label{sec:metho}
\begin{figure}[!ht]
		\centering
		\includegraphics[width=0.8\textwidth]{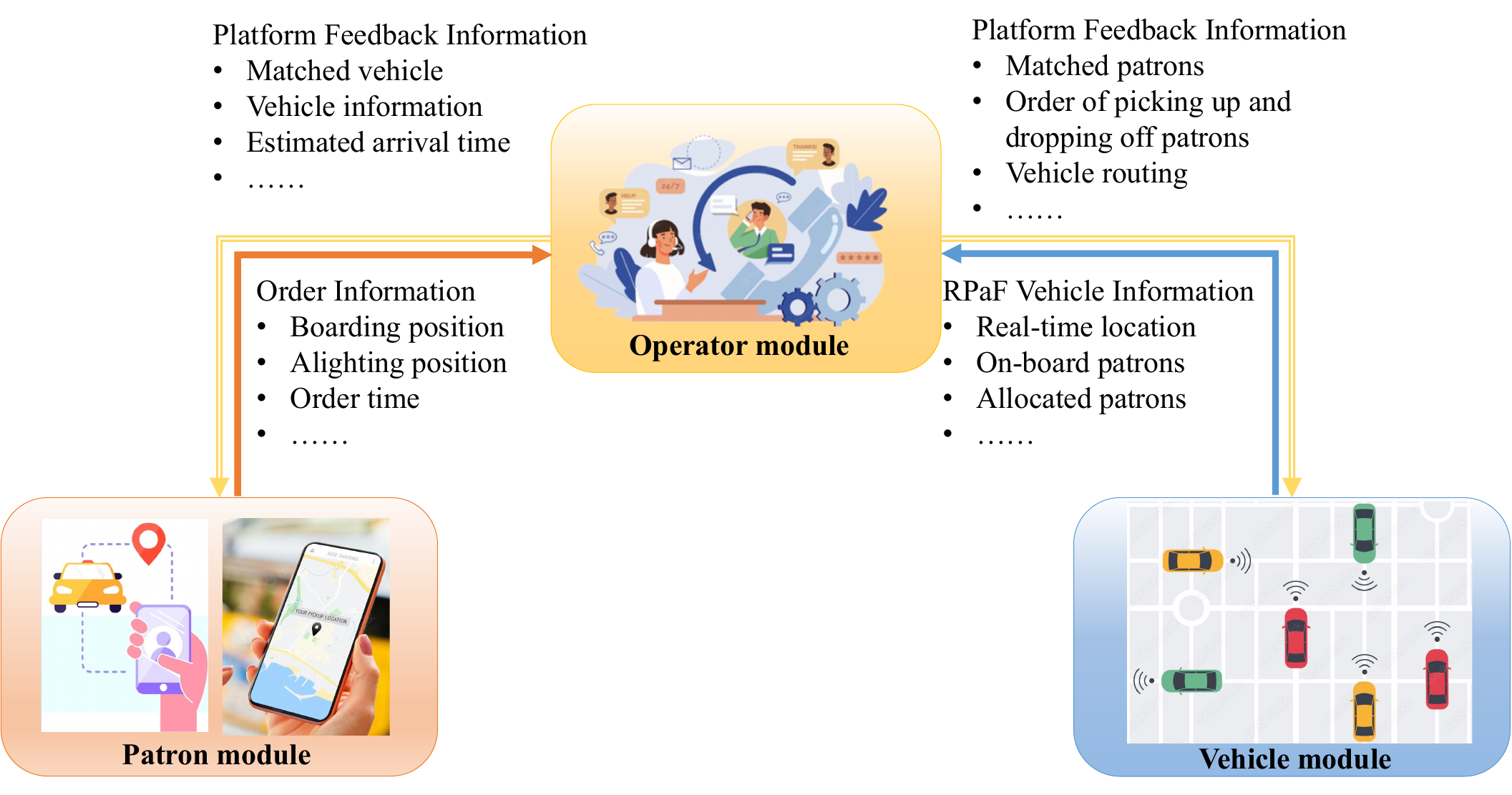}
		\caption{Interaction among three simulation modules. (Source of graphic illustration: \url{https://www.freepik.com})}\label{fig:sti-frame}
    \end{figure}

The simulation framework contains three simulation modules for the operator, patrons, and vehicles.
We utilize Simulation of Urban MObility (SUMO) as the tool to develop a customized time-based simulation environment for RPaF services.\footnote{{The simulation platform has been open-sourced at \url{https://github.com/yxt19981119/SUMO-RPaF.git}}.} SUMO is an open-source microscopic simulation software widely used in transportation studies \citep{Huang2020, Zhu2020TRR}. Specifically, we use the Traci (Traffic Control Interface) of SUMO to manipulate the actions of simulated entities, i.e., the operator, patrons, and vehicles. Accordingly, we devise three interactive modules, as shown in Figure \ref{fig:sti-frame}, which are described in the following subsections. 
	
\subsection{Operator module}
\begin{figure}[!ht]
		\centering
		\includegraphics[width=\textwidth]{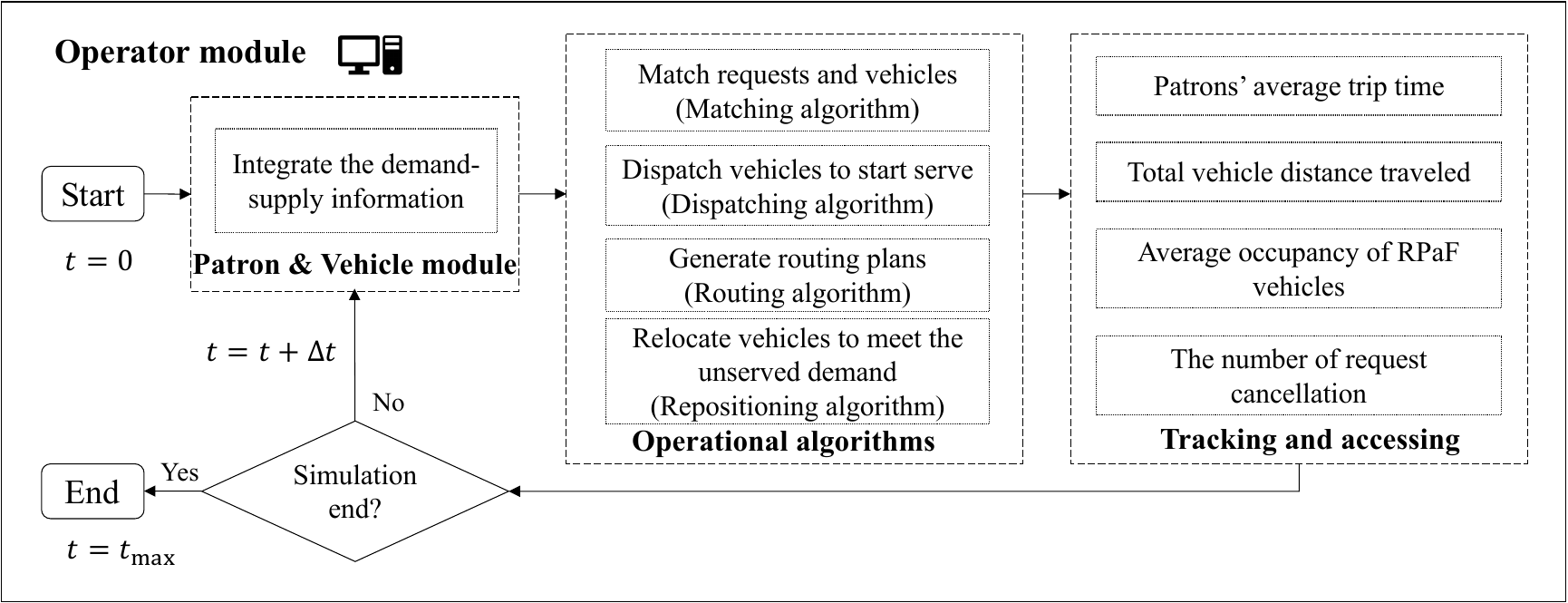}
		\caption{Simulation flow chart of the {operator module (where $\Delta t$ is simulation timestamp)}.}\label{fig:operator}
	\end{figure}

The operator module is the core of the simulation. It employs the above-proposed operational algorithms to control or impact the actions of the other two modules, as illustrated in Figure \ref{fig:operator}. Specifically, the operator module dynamically (i) matches requests and vehicles (called the ``matching algorithm''), (ii) dispatches vehicles to start transport service (``dispatching algorithm''), (iii) generates routing plans (``routing algorithm''), and (iv) relocates vehicles to meet the unserved demand (``repositioning algorithm''). 

Note that each of the four types of algorithms may have several alternatives that have been proposed in the literature \citep{WANG2019122}. Our module-based simulation environment can accommodate any specific algorithms of interest for study. 

 The operator module is also accountable for assessing the system's performance. Based on information input from the other two modules, it computes metrics of interest, including patrons' average trip time, total vehicle distance traveled, average occupancy of RPaF vehicles, etc. 
	
\subsection{Patron module}
Patrons in this module are indexed by $i=1,2,...$ following the emergence order of their requests, and each is associated with {a three-tuple state $r_i = \langle t_i, \Vec{l}_i,\delta_i \rangle$, which are being updated at each simulation timestamp.} Here, {$t_i$} represents patron $i$'s request call-in timestamp, and vector {$\Vec{l}_i$} denotes her origin or destination location. For instance, $\Vec{l}_i = (0,0)$ marks the connection point between the suburb and the freeway segment, as shown in Figure \ref{fig:layout}. The $\delta_i$ is an indicator variable, $\delta_i=0$ by default, meaning no match with any vehicles, $\delta_i=1$ if matched with a vehicle, $\delta_i=2$ if picked up by the designated vehicle, and $\delta_i = 3$ if finished her trip. 

\begin{figure}[!ht]
	\centering
 	\includegraphics[width=\textwidth]{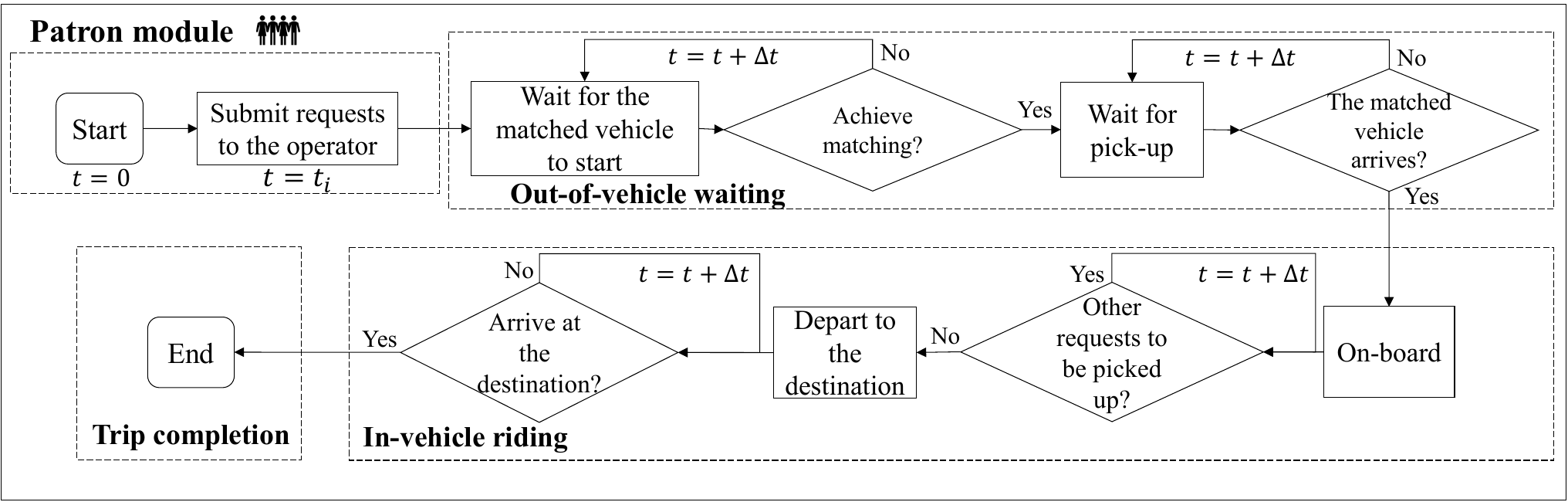}
	\caption{Simulation flow chart of patron module (where $\Delta t$ is simulation timestamp). 
	}\label{fig:patron}
	\end{figure}

The patron module simulates patrons' behaviors in four stages, as shown in Figure \ref{fig:patron}:
    \begin{enumerate}[label={(\arabic*) },ref=\arabic*]
	    \item Trip generation: Patrons are generated following specific temporal and spatial distributions.   
	    \item Out-of-vehicle waiting: Patrons are put on wait at their origins (e.g., residential homes or the hub) until boarding. 
	    \item In-vehicle riding: If a patron travels outbound, she first rides the vehicle to pick up other matched requests, if any are still waiting. And then, they ride together with the vehicle directly to the hub without any stops. The process is reversed for inbound patrons, who ride the vehicle first in a line-haul segment and then a tour for drop-offs. Note that the tour segments of pick-ups and drop-offs are separated to ensure patron acceptance.
	    \item Trip completion: Once patrons arrive at their destinations (the hub or residential places),  they complete their trips and exit the simulation. 
	\end{enumerate}
	
Patron $i$'s total trip time, {$T_i$} [seconds], is recorded as the sum of the out-of-vehicle waiting time ($T^w_i$) and the in-vehicle riding time ($T^r_i$).


\subsection{Vehicle module} \label{sec_vehicle_module}
We consider a fleet of $N$ identical vehicles, indexed by $n=1,..., N$, with capacity $C$ [seats/vehicle]. Each vehicle is attached with a four-tuple state $v_n = \langle o_n, \Vec{l}_n, X_n, \delta_n \rangle $. Here, $o_n$ denotes the vehicle occupancy (i.e., the number of on-board patrons); {$\Vec{l}_n$} represents the location of the vehicle; $X_n = \left\{i |x_{n, i}=1 \& \delta_i =1, \right\}$ is the set of requests assigned to the vehicle $n$, where $x_{n, i}=1$ indicates a match between vehicle $n$ and patron $i$ and $x_{n, i}=0$ otherwise; and $\delta_n$ here indicates outbound service ($\delta_n=\text{`OB'}$) or inbound service ($\delta_n=\text{`IB'}$).

\begin{figure}[!ht]
		\centering
 		\includegraphics[width=\textwidth]{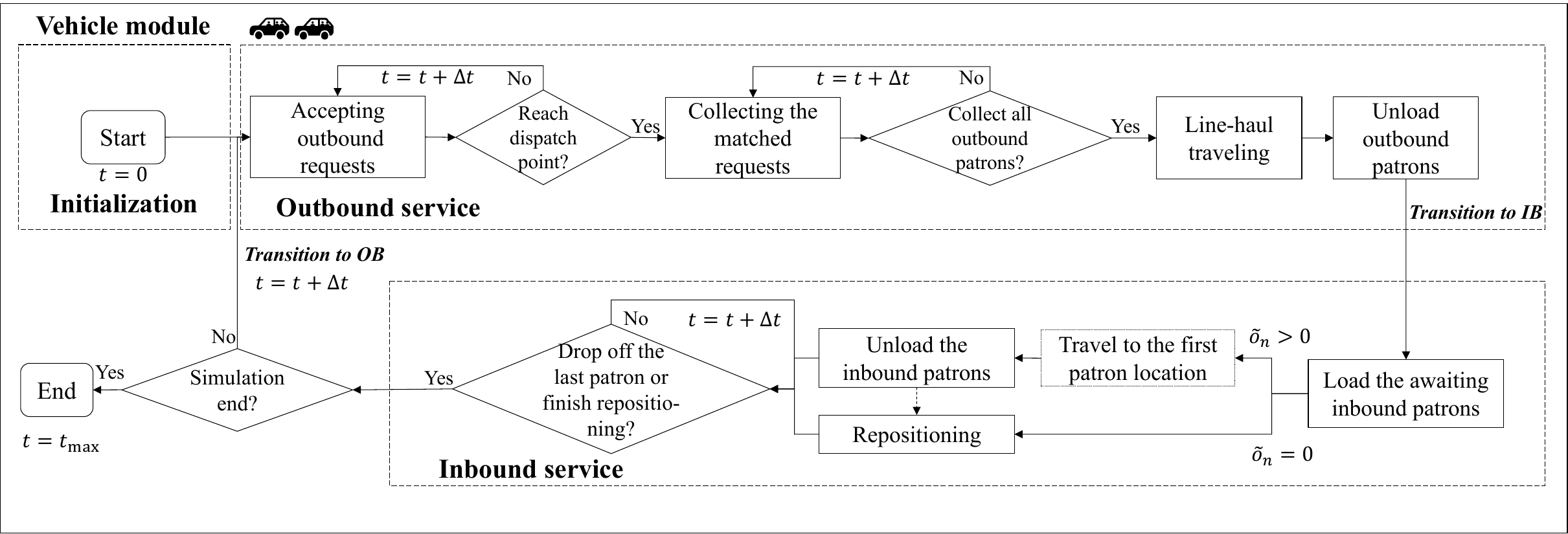}
		\caption{Simulation flow chart of vehicle module (where $\Delta t$ is simulation timestamp). 
		}\label{fig:RPVs}
	\end{figure}

This module simulates the circulation of vehicles in bidirectional trips, as shown in Figure \ref{fig:RPVs}. 
	\begin{enumerate}[label={(\arabic*) },ref=\arabic*]
    	    \item Outbound trip:
    	    \begin{enumerate}[label=(\arabic{enumi}.\arabic*)]
    	        \item Requests-accepting: Vehicles with $o_n = 0$ and $|X_n| < u$ keep accepting request assignments from {the matching algorithm (operator module)}. Each assignment updates the state of the matched patron with $\delta_i = 1 $ and adds the new matched request into $X_n$. 
    	        \item Requests-collecting: Vehicles are dispatched by the {dispatching algorithm (operator module)} to pick up the matched requests in $X_n$. They follow the route planned by {the routing algorithm (operator module).} Upon each pick-up ($\Vec{l}_n = \Vec{l}_i, i \in X_n$), the vehicle's state is updated with an increment of $o_n+1$, and the patron' state is also updated with $\delta_i = 2$.
    	        \item Line-haul traveling: After collecting all matched requests (i.e., $o_n = u$ or $\delta_i = 2, \forall i \in X_n$), the vehicle transports them directly to the hub without stops.
    	        \item Hub-unloading: Upon arrival at the hub, the outbound vehicle unloads all patrons ($\delta_i = 3, \forall i \in X_n$) and then transitions to the inbound-service state ($\delta_n=\text{IB}$). 
    	    \end{enumerate}
    	    
    	    \item Inbound trip: 
    	    
    	    \begin{enumerate}[label=(\arabic{enumi}.\arabic*)]
    	        \item Hub-loading: Inbound vehicles, {assumed to follow a ``stop-and-go'' rule}\footnote{{Under this rule, RPaF vehicles are prohibited from long-time parking to conserve valuable space at the hub. A similar concept is the commonly observed ``kiss-and-ride'' at metro stations or local rail stations.}}, load the waiting inbound patrons, if any, in a First-In-First-Out (FIFO) order. The vehicle state is changed to $\langle \tilde{o}_n, \Vec{l}_n, X_n,\text{IB} \rangle$, where $\tilde{o}_n \in [0, C]$ is the number of on-board inbound patrons, ensuring it does not exceed the vehicle capacity $C$ [seats/vehicle]; and the set $X_n$ contains inbound patrons onboard vehicle $n$. 
    	        \item Line-haul traveling: If the number of onboard patrons $\tilde{o}_n>0$, the vehicle begins traveling directly to the first location in the tour made by the routing algorithm. Otherwise if $\tilde{o}_n=0$, the vehicle returns to {the assigned location by the repositioning algorithm.}
    	        \item Requests-delivering: The vehicle unloads patrons in $X_n$ following the plan by the routing algorithm. After dropping off the last patron, the vehicle, {if not relocated}, changes to the outbound-service state with $\delta_n = \text{OB}$; {otherwise, it changes the state upon arrival to the repositioned location.}
    	    \end{enumerate}
    \end{enumerate}
    
\section{Experimental studies}\label{sec:experi}
\subsection{Set-up}
{The experimental simulation's baseline setting is as follows. The suburban area spans 5 km $\times$ 5 km, with a grid street network spaced by {0.1 km}\footnote{{This is set according to the average city block in the US, which is around 311 feet or 95 meters \citep{daganzo2019public}.}} and a freeway segment of length $L=5$ km. {All streets are two-way and have the same layout, e.g., three lanes in each direction. It is assumed that all intersections cause a standard delay of 10 seconds for vehicles passing through or making turns. Turning movements (including U-turns) can only be made at intersections.} The RPaF vehicles are sedans with a maximum capacity of $C=4$ [seats/vehicle]. After a few rounds of tests, we set the initial fleet size to be {$N=27$} [vehicles], so that at system steady states, the supply is generally balanced with the following demand.} 

{We are particularly interested in the morning commute peak period when the demand is typically more aggregated than other periods and thus more crucial for the operation study. The total demand is set to be 180 and 20 patrons per hour for outbound and inbound directions, respectively. The corresponding densities in uniform demand distributions are $\lambda_{\rm OB}=7.2$ and $\lambda_{\rm IB}=0.8$ [patrons/sq. km/hour]. The emergence of requests follows Poisson processes. 

{We further consider non-uniform demand distributions using the following distance-decaying functions \citep{fan2024optimal},}
    \begin{align} \label{eq_hete}
        \lambda_{\rm OB}(\Vec{l})=\lambda_{\rm OB}\text{exp}(-\mu_{\rm OB} \|\Vec{l}\|),\qquad
        \lambda_{\rm IB}(\Vec{l})=\lambda_{\rm IB}\text{exp}(-\mu_{\rm IB} \|\Vec{l}\|)
    \end{align}
{where $\Vec{l}$ marks a location in the suburban region and $\|\Vec{l}\|$ yields the distance to the freeway; and parameters $\mu_{\rm OB},\mu_{\rm IB}$ reflect the demand sensitivity to the distance, as shown in Figure \ref{fig:hete_decay}, where they are set to be $\mu_{\rm OB} = \mu_{\rm IB} = 0.1$.} {The $\lambda_{\rm OB}$ and $\lambda_{\rm IB}$ in Eq. (\ref{eq_hete}) reflect the latent demand derived from the socio-economic characteristics of the suburb. Therefore, the above non-uniform distributions illustrate the influence of distance on actual demand, which has been noted in feeder systems \citep[e.g.,][]{fan2024optimal, fan2018joint, zhen2023feeder}.}

    \begin{figure}[!ht]
		\centering
		\includegraphics[width=0.65\textwidth]{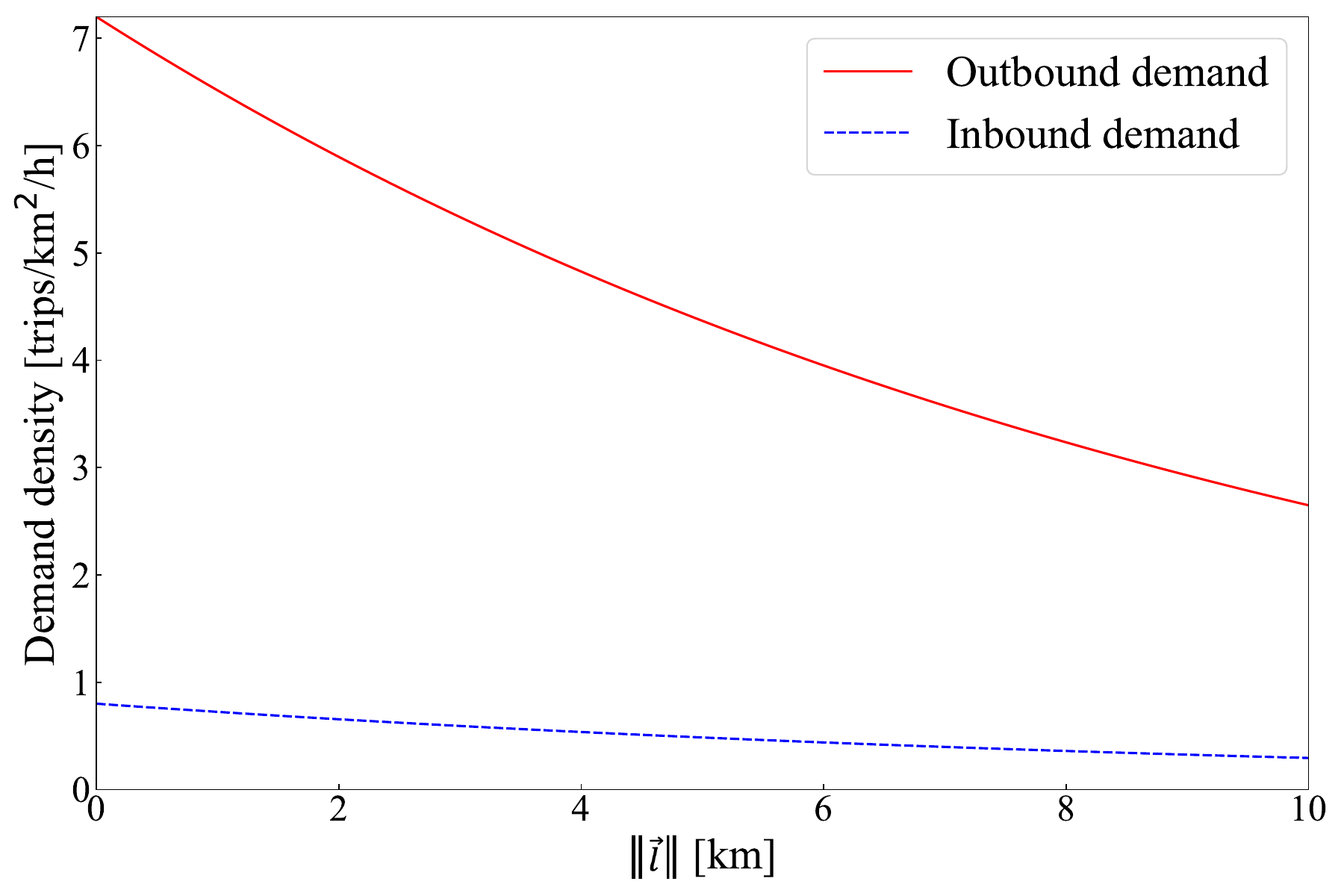}
		\caption{Non-uniform demand pattern (with $\mu_{\rm OB} = \mu_{\rm IB} = 0.1$).}\label{fig:hete_decay}
    \end{figure}

{We begin by setting the occupancy target as $u = C = 4$ and accordingly obtain the matching buffer distance $\Delta^* = 1.67$ km for uniform demand scenarios by Eq. (\ref{optimal_Delta}) in the Manhattan metric (since the street network is a grid network). As for non-uniform demand scenarios, the $\Delta^*$ is computed for each vehicle according to the local demand density.}

The simulation runs for 2.5 hours, with a timestamp interval of $\Delta t = 1$ second. The first 30 minutes serve as a warm-up period. Other parameter values are summarized in Table \ref{simulation:parameter}, which are adopted from previous studies (e.g., \citealp{EDWARDKIM201967, daganzo2019public}).}
    \begin{table}[ht]
		\caption{\textbf{Simulation parameter values.}}\label{simulation:parameter}
 		\vspace{0.1in}
		\centering
		\begin{tabular}{c c c c }
			\hline
			\textbf{Parameter} & \textbf{Baseline Value} & \textbf{Parameter} & \textbf{Baseline Value} \\
			\hline
			$S,S^{'}$ [km/hour] & $60,30$  & $t_{d}$ [seconds/stop] & 3  \\
			
			$\beta$ [\$/hour] & 20  & $c_f$ [\$/dispatch] & 18  \\
			
            $\alpha $ & 0.5 & {$\Bar{\tau}$ [hour]} & {0.1} \\
            \hline
		\end{tabular}
	\end{table}
 
{For completeness, our simulation platform can also account for general trips that start and end within the suburb. To this end, we develop an algorithm based on \cite{daganzo2019public}'s Ride-Matching (RM) concept, which pools requests by matching their origins and destinations. Details are given in \ref{app_RM}. Note that the operational strategy of RM is fundamentally different from that of feeder services and should be independently operated with a dedicated fleet. {Considering demand fluctuations may enhance operations with a shared fleet for both feeder and general services, but this is beyond the scope of this paper.} Therefore, the subsequent analysis will exclusively focus on feeder services without explicitly exploring the general-trip service.}

\subsection{Evaluation of RPaF services}
This section evaluates the performance of RPaF services. We begin by analyzing the impact of {the zoning strategy (Section \ref{sec_zoning}), {which is a common strategy in feeder services \citep{fan2024optimal, EDWARDKIM201967}. Although the proposed algorithms work with or without zoning, we seek to understand its impact on overall system performance of the proposed RPaF.} We then examine the effectiveness of the matching algorithm (Section \ref{sec_matching}), the dispatching algorithm (Section \ref{sec_dispatching}), and finally the repositioning algorithm (Section \ref{sec_repositioning}). The routing algorithm is not evaluated in detail as it is an implementation of the traditional TSP in specific street networks. Exact solutions to vehicle routing problems are typically found within 0.1 seconds for all simulated scenarios. For each evaluation, the performance metrics take an average of 50 runs with the same sequence of generated demand.}

\subsubsection{Zoning strategy} \label{sec_zoning}
\begin{figure}[!ht]
		\centering
		\subfigure[]{\includegraphics[width=0.48\textwidth]{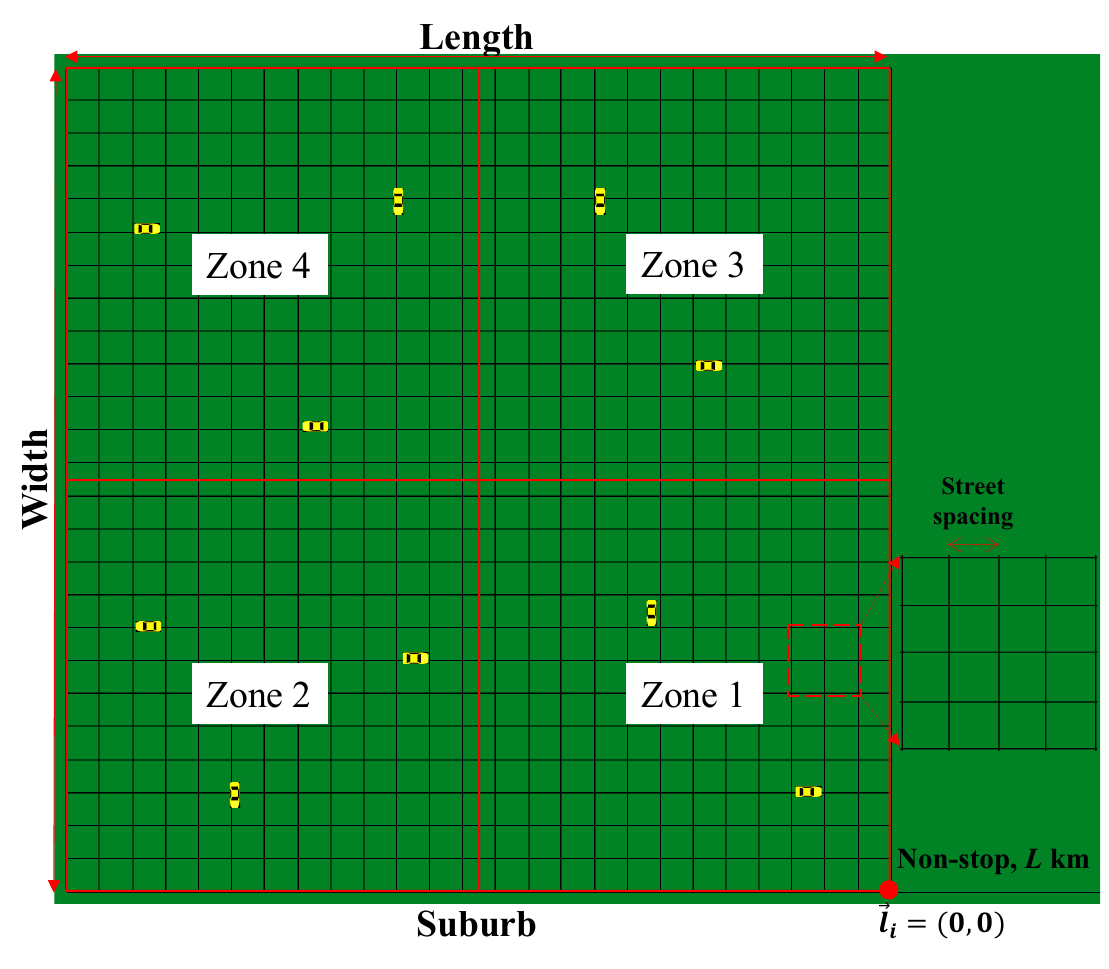}\label{fig:sub_zone1}}
        \subfigure[]{\includegraphics[width=0.48\textwidth]{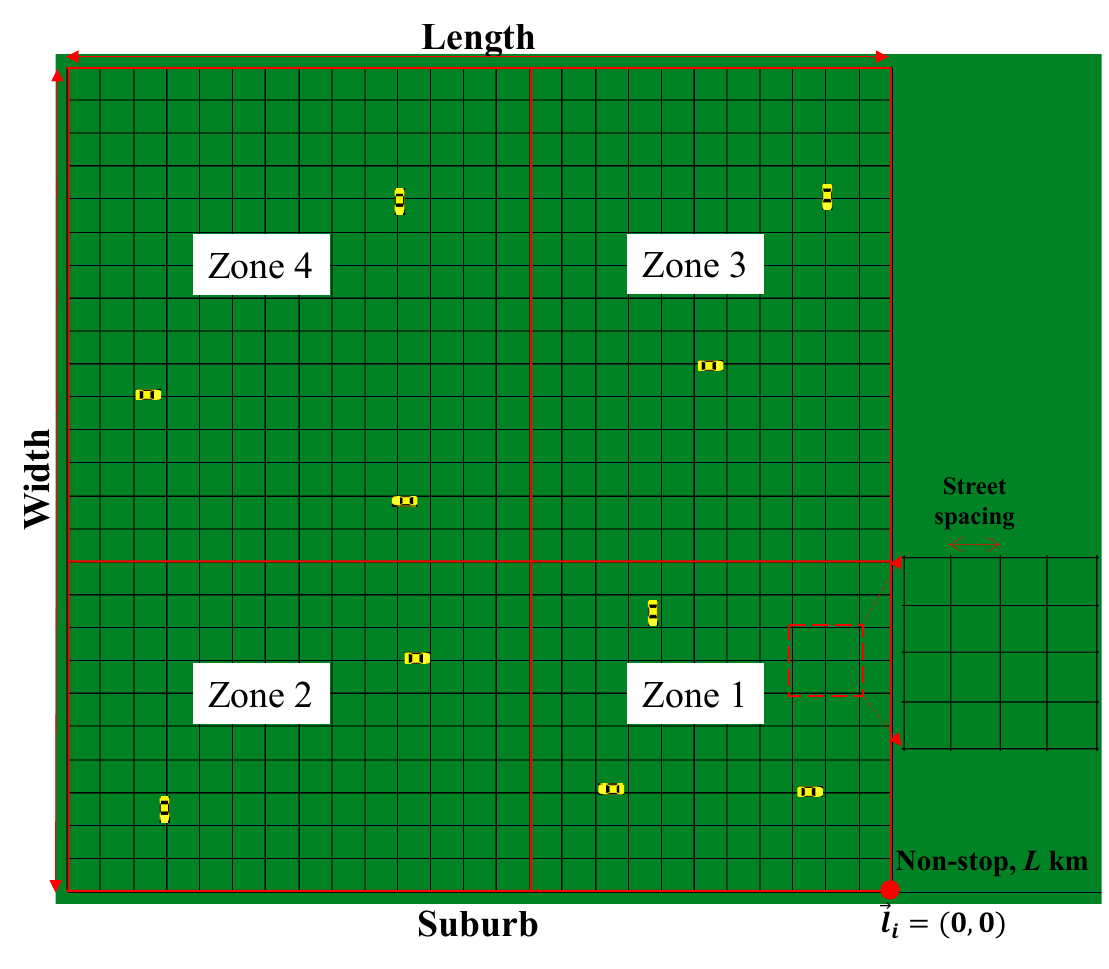}\label{fig:sub_zone2}}
		\caption{Zoning schemes under (a) uniform and (b) non-uniform demand scenarios.}\label{fig:sub_zone}
    \end{figure}

{The zoning strategy partitions the entire service region into multiple zones and designates a fleet of vehicles serving specifically for each zone.\footnote{The zoning strategy implicitly assumes that vehicles and patrons at the hub can be organized according to their respective zones. This requires infrastructure like sufficient parking, waiting spaces, or designated channels.} It has proven effective in RPaF services \citep{fan2024optimal} and also in Flex-FBT systems \citep[e.g.,][]{EDWARDKIM201967}. In contrast to those analytical studies based on idealized networks, we examine the effectiveness of zoning by simulations in specific street networks. However, we intend not to find the best partitions, which would be challenging and require further research (e.g., using graph-based network optimization).} {Instead, we divide the suburban region into four zones, as depicted in Figure \ref{fig:sub_zone1} and \ref{fig:sub_zone2}. These zones are identical when facing uniform demand scenarios (Figure \ref{fig:sub_zone1}), but they differ when dealing with non-uniform demand scenarios (Figure \ref{fig:sub_zone2}). Inspired by \cite{fan2024optimal}, Zone 1/4 in Figure \ref{fig:sub_zone2} is assigned the smallest/largest size due to the highest/lowest average demand density. Subsequently, we distribute the fleet of vehicles in proportion to the demand in each zone. Note that the results obtained from the current zone partition are somewhat conservative, and any enhancement in the zone partition could further improve the system's performance.\footnote{According to the finding in \cite{fan2024optimal}, zone size has a mild impact on the performance of the zoning strategy in RPaF services, particularly under low level of inbound demand.}}

{Table \ref{tab:zone} presents the performance of RFaF with and without zoning strategy under the baseline setting, showing the service rate and average trip time components for patrons. Here, the service rate refers to the percentage of patrons who have not canceled their requests and completed their trips in the feeder system. The findings demonstrate that the zoning strategy improves the service rate and decreases the trip time for patrons by up to 9.8\%. However, the time savings are less apparent in non-uniform demand scenarios because most requests aggregate in zone 1 with short trip distances.}

\begin{table}[!ht]
    \centering
\begin{threeparttable}
    \caption{{Performance of RPaF with and without zoning under the baseline scenario.}}
    \label{tab:zone}
    \begin{tabular}{c|c|c|c|c|c}
         \hline
         \makecell{\textbf{Demand} \\ \textbf{distribution}} & \makecell{\textbf{Zoning}} & \makecell{\textbf{Service} \\ \textbf{rate$^*$ [\%]}}
         & \makecell{\textbf{Waiting} \\ \textbf{time [h]}} & \makecell{\textbf{In-vehicle} \\ \textbf{time [h]}} & \makecell{\textbf{Trip} \\ \textbf{time [h]}} \\
         \hline
         \multirow{2}{*}{Uniform} & With & 91 & 0.12 & 0.25 & 0.37 \\
         & Without & 90 & 0.15 & 0.26 & 0.41 \\
         \hline
         \multirow{2}{*}{Non-uniform} & With & 93 & 0.110 & 0.207 & 0.317 \\
         & Without &91.5 & 0.113 & 0.210 & 0.323 \\
         \hline
    \end{tabular}
    \begin{tablenotes}  
        \footnotesize  
        \item $^*$ The service rate refers to the percentage of patrons who completed their trips without canceling requests.  
    \end{tablenotes}  
\end{threeparttable}
\end{table}

\begin{figure}[!ht] 
		\centering
		\subfigure[Uniform demand scenarios]{\includegraphics[width=0.48\textwidth]{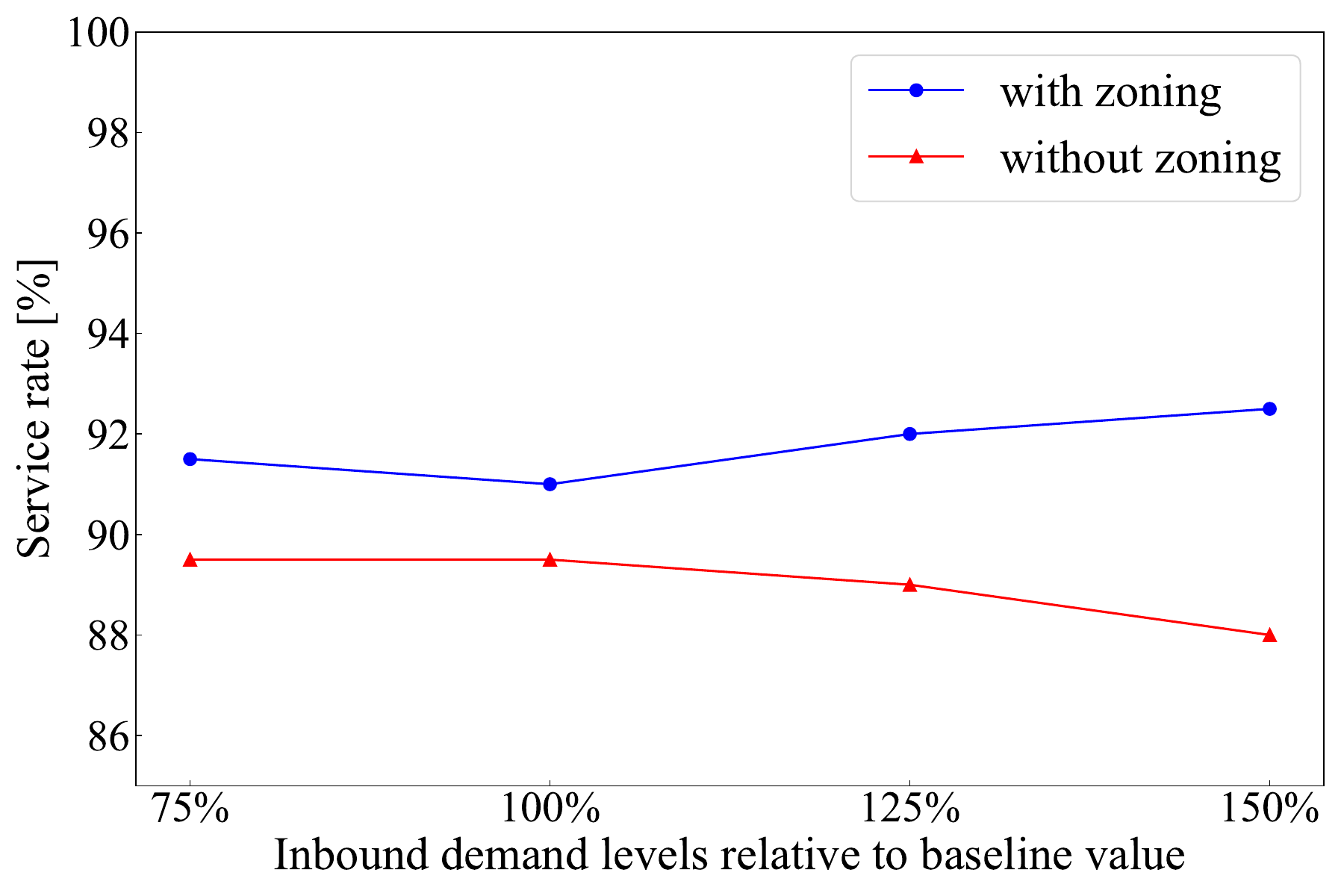}\label{fig:match_uniform}}
        \subfigure[Non-uniform demand scenarios]{\includegraphics[width=0.48\textwidth]{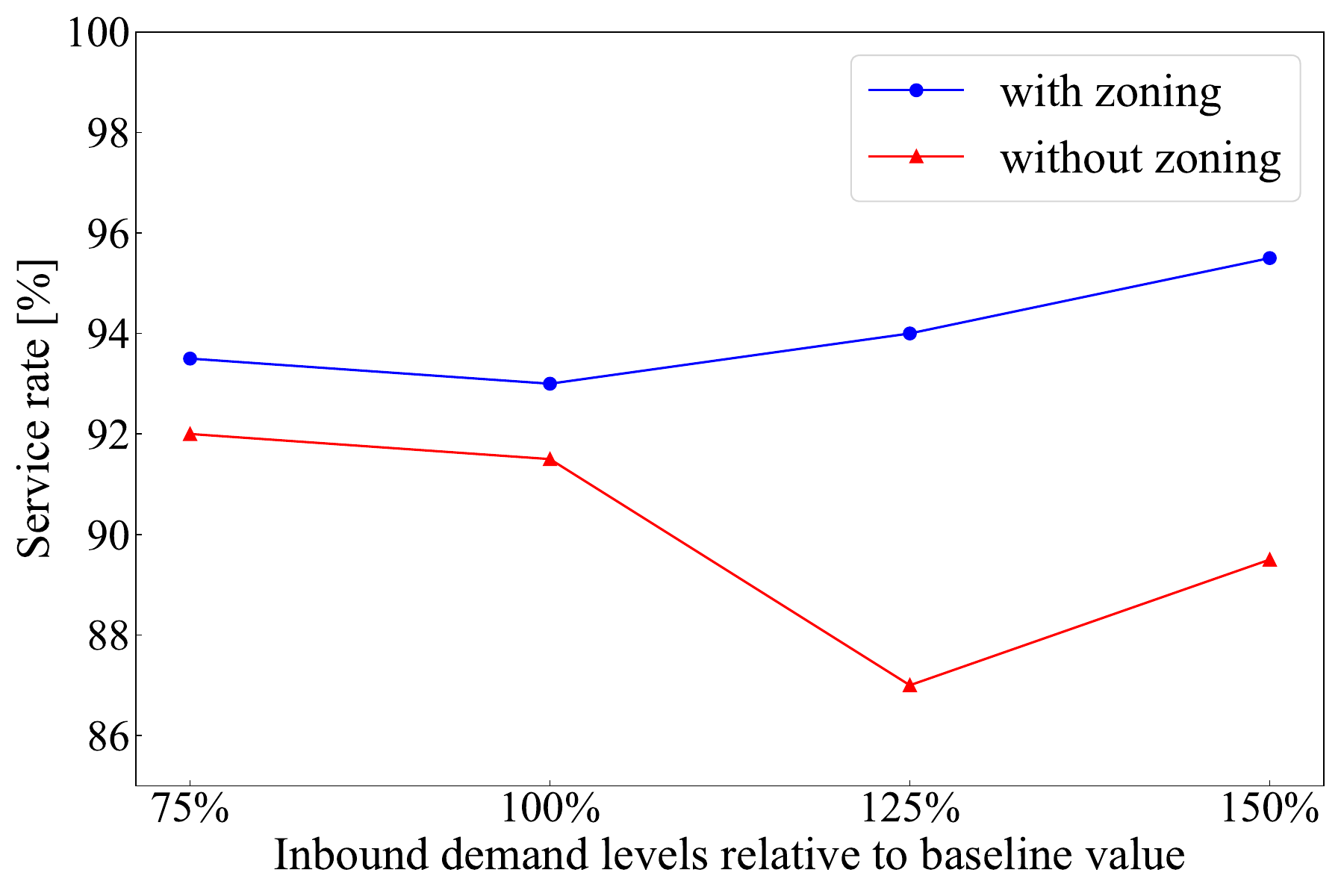}\label{fig:match_non_uniform}}
		\caption{Service rates in RPaF with and without zoning.}\label{fig:zonging_service}
  \label{fig_service_rate_changes}
    \end{figure}

\begin{figure}[!h]
        \centering
        \includegraphics[width=0.6\textwidth]{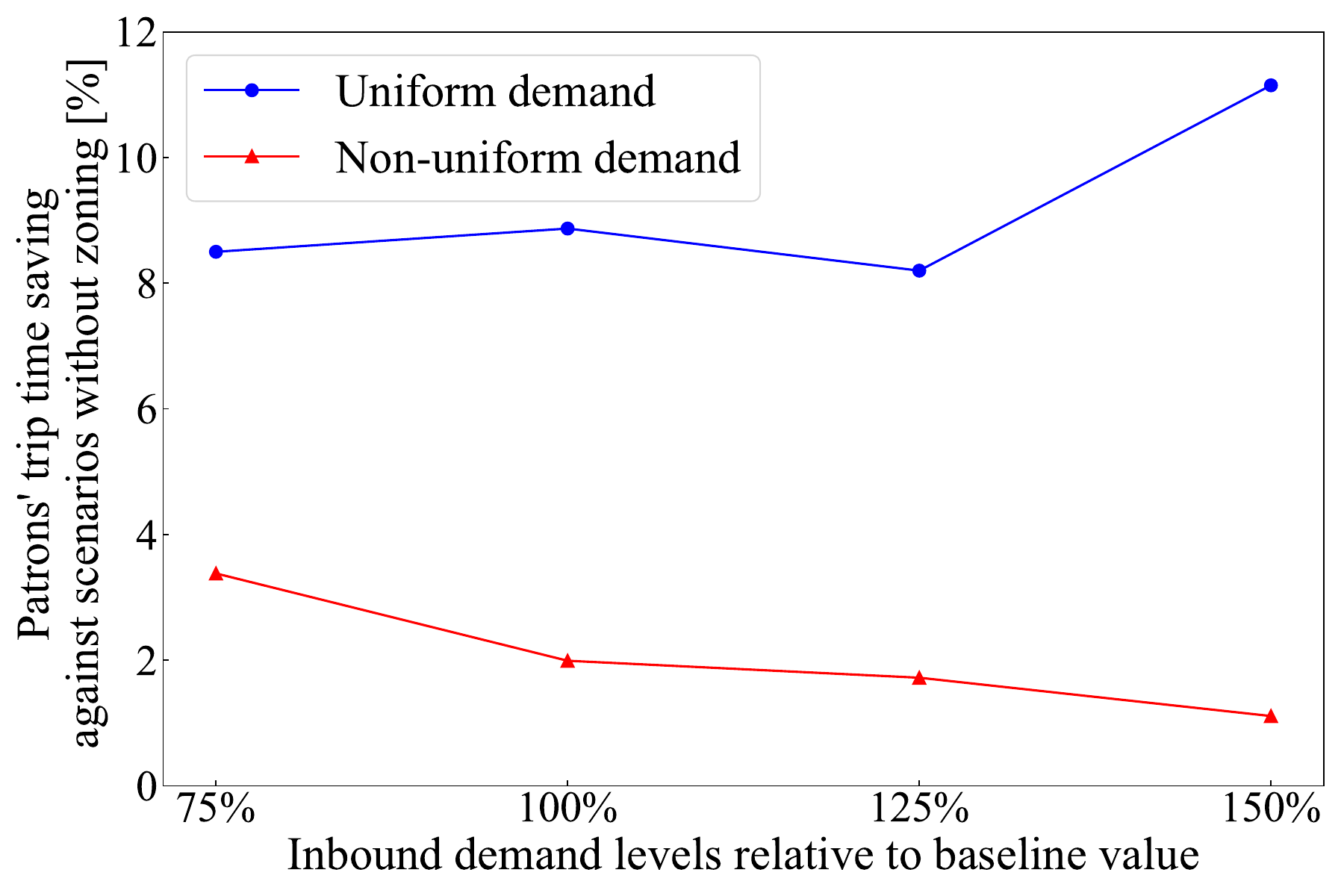}
        \caption{Patrons' trip time savings against scenarios without zoning.}
        \label{fig:zone}
    \end{figure}

{We further conducted sensitivity analyses regarding inbound demand levels by scaling the baseline $\lambda_{\rm IB}$.} The outbound demand is not chosen for analysis because it is theoretically proven to be irrelevant to zoning strategies in RPaF services \citep{fan2024optimal}. {Figures \ref{fig_service_rate_changes} and \ref{fig:zone} illustrate the changes in service rates and patrons' trip time savings. Specifically, Figures \ref{fig_service_rate_changes}(a) and (b) show that higher demand levels can lead to increased service rates for RPaF with zoning, while RPaF without zoning will have lower service rates. Figure \ref{fig:zone} indicates that patrons can achieve trip time savings over 11\% in uniform demand scenarios, while non-uniform demand shows observable savings (e.g., over 3\%) at lower demand levels.}

\begin{figure}
        \centering
        \includegraphics[width=0.6\linewidth]{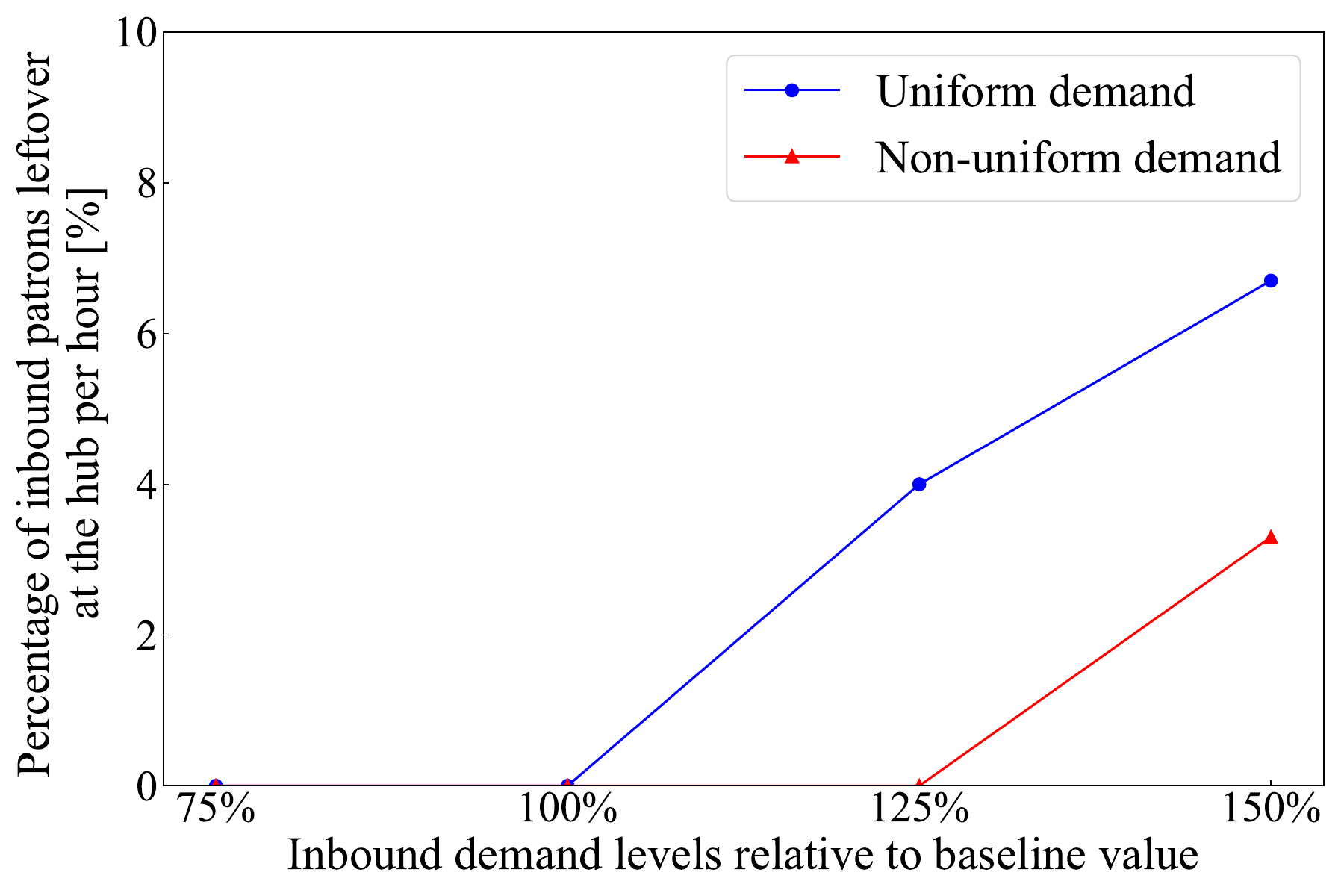}
        \caption{{Percentage of inbound patrons leftover by the first departing vehicle at the hub per hour.}}
        \label{fig:zoning_extra}
    \end{figure}

{Nonetheless, we notice that as the inbound demand levels increase, some patrons at the hub may be unable to board the first departing vehicle due to the capacity limit and must continue waiting for the next one. We record the simulation instances of these patrons experiencing additional wait per hour and present the percentages as shown in Figure \ref{fig:zoning_extra}. The results reveal that while the percentage of leftover patrons is negligible at low demand levels, it steadily rises with increasing demand, remaining below 10\% even at 1.5 times the baseline demand. In all simulation scenarios, we observe no patrons waiting for more than two vehicles to depart from the hub. In other words, all inbound patrons board either the first or second departing vehicle. It is worth noting that the above results are dictated by the supply of RPaF vehicles, which may mitigate the issue with a larger fleet size or vice versa.}

{The zoning strategy remains in the following scenarios due to its benefits demonstrated above.}

\subsubsection{Matching algorithm}\label{sec_matching}
{This section examines the effectiveness of the batch-based matching algorithm by comparing scenarios with and without the optimized matching buffer distance $\Delta^*$. For the latter, the (outbound) requests are always matched to the closest available vehicle within the same zone and no batch concept is employed.}

\begin{figure}[!ht]
		\centering
		\subfigure[Uniform demand scenarios]{\includegraphics[width=0.48\textwidth]{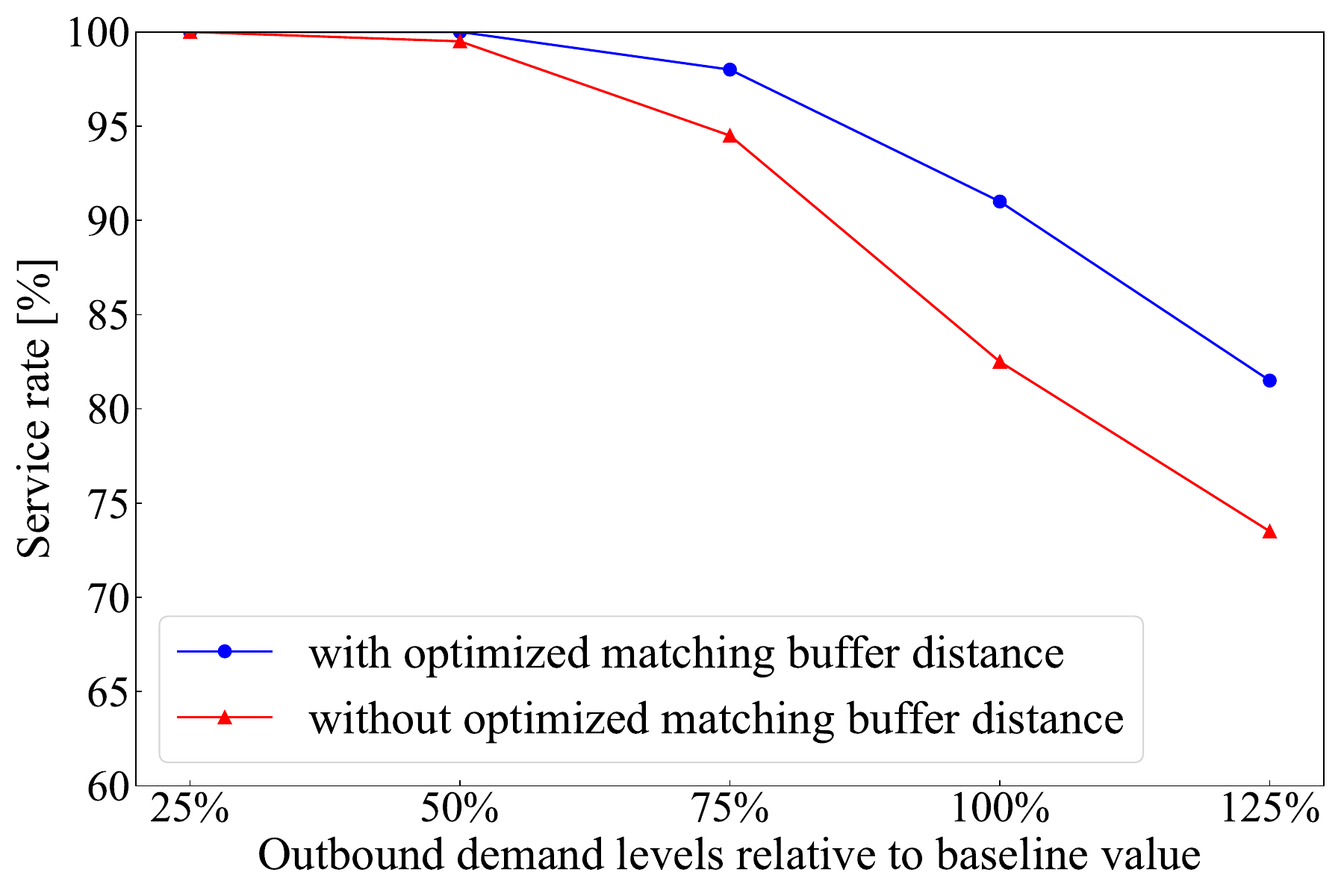}\label{fig:match_service_uniform}}
        \subfigure[Non-uniform demand scenarios]{\includegraphics[width=0.48\textwidth]{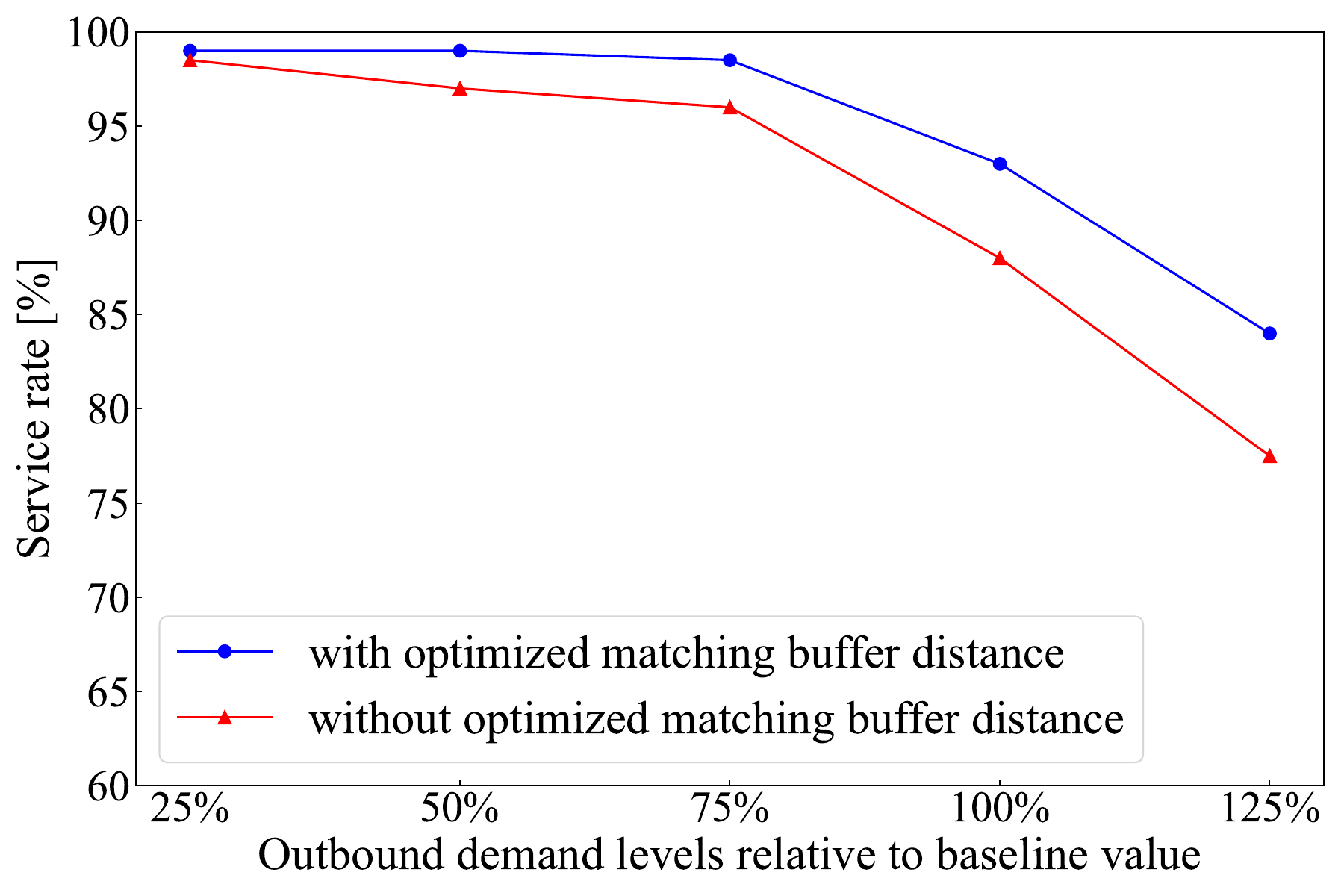}\label{fig:match_service_non_uniform}}
		\caption{{Service rates of RPaF with and without optimized matching buffer distance.}}\label{fig:match_service_result}
    \end{figure}

{Figures \ref{fig:match_service_result}(a) and (b) depict service rates of RPaF across various outbound demand levels. It is evident that in both uniform and non-uniform demand scenarios, the optimized buffer distance helps RPaF enhance service rates, although the percentage declines as demand levels increase.}

{To enable a fair comparison, we increase the fleet size of the counterpart system to match the proposed system's service rate of 90\%. Table \ref{tab:match} summarizes the results under the baseline scenario, showing that the counterpart requires 14.8\% and 12\% more fleet, while patrons' average trip time remains similar between the two systems.}

\begin{table}[!ht]
    \centering
    \caption{{Performance of RPaF with and without optimized matching buffer distance under the baseline scenario.}}
    \label{tab:match}
    \begin{tabular}{c|c|c|c|c|c}
         \hline
         \makecell {\textbf{Demand} \\ \textbf{distribution}} & \makecell{\textbf{Optimized matching}\\ \textbf{buffer distance}}
         & \makecell{\textbf{Waiting} \\ \textbf{time [h]}} & \makecell{\textbf{In-vehicle} \\ \textbf{time [h]}} & \makecell{\textbf{Trip} \\ \textbf{time [h]}}& \makecell{\textbf{Fleet size} \\ \textbf{[vehicles]}} \\
         \hline
         \multirow{2}{*}{Uniform} & With & 0.12 & 0.25 & 0.37 & 27\\
         & Without & 0.13 & 0.24 & 0.37 & 31\\
         \hline
         \multirow{2}{*}{Non-uniform} & With & 0.12 & 0.21 & 0.33 & 25\\
         & Without & 0.13 & 0.20 & 0.33 & 28\\
         \hline
    \end{tabular}
\end{table}

\subsubsection{Dispatching algorithm}\label{sec_dispatching}
The dispatching algorithms depend on the occupancy target $u$. This section examines hard- and soft-target strategies with $u \in \left\{2,3,4\right\}$. {Table \ref{tab:dispatch} shows the baseline scenario results, indicating that higher occupancy targets generally improve RPaF service rates for both dispatch algorithms. Furthermore, given the same occupancy targets, the hard-target algorithm achieves higher service rates and shorter average trip times for patrons compared to the alternative algorithm.}
\begin{table}[!ht]
    \centering
    \caption{{Performance of RPaF with hard- and soft-target dispatching algorithms under the baseline scenario.}}
    \label{tab:dispatch}
    \begin{tabular}{c|c|c|c|c|c|c}
         \hline
         \makecell{\textbf{Demand} \\ \textbf{distribution}} & \makecell{\textbf{Dispatching} \\ \textbf{algorithm}} & \makecell{\textbf{Occupancy} \\ \textbf{target $u$}} & \makecell{\textbf{Service} \\ \textbf{rate [\%]}}
         & \makecell{\textbf{Waiting} \\ \textbf{time [h]}} & \makecell{\textbf{In-vehicle} \\ \textbf{time [h]}} & \makecell{\textbf{Trip} \\ \textbf{time [h]}} \\
         \hline
         \multirow{6}{*}{Uniform} & \multirow{3}{*}{Hard-target} & 4  & 91 & 0.12 & 0.25 & 0.37\\
         &  & 3 & 85.5 & 0.11 & 0.24 & 0.35 \\
         &  & 2 & 68 & 0.12 & 0.24 & 0.36 \\
         \cline{2-7}
         & \multirow{3}{*}{Soft-target} & 4 & 79.5 & 0.13 & 0.25 & 0.38 \\
         &  & 3 & 77 & 0.12 & 0.24 & 0.36 \\
         &  & 2 & 64.5 & 0.12 & 0.24 & 0.36 \\
         \hline
         \multirow{6}{*}{Non-uniform} & \multirow{3}{*}{Hard-target} & 4 & 93 & 0.12 & 0.21 & 0.33 \\
         &  & 3 & 86.5 & 0.12 & 0.20 & 0.32 \\
         &  & 2 & 73 & 0.11 & 0.19 & 0.30 \\
         \cline{2-7}
         & \multirow{3}{*}{Soft-target} & 4 & 82.5 & 0.12 & 0.21 & 0.33 \\
         &  & 3 & 83 & 0.12 & 0.21 & 0.33 \\
         &  & 2 & 70.5 & 0.11 & 0.20 & 0.31 \\
         \hline
    \end{tabular}
    
\end{table}

\begin{figure}[!ht]
		\centering
		\subfigure[Uniform demand scenarios]{\includegraphics[width=0.48\textwidth]{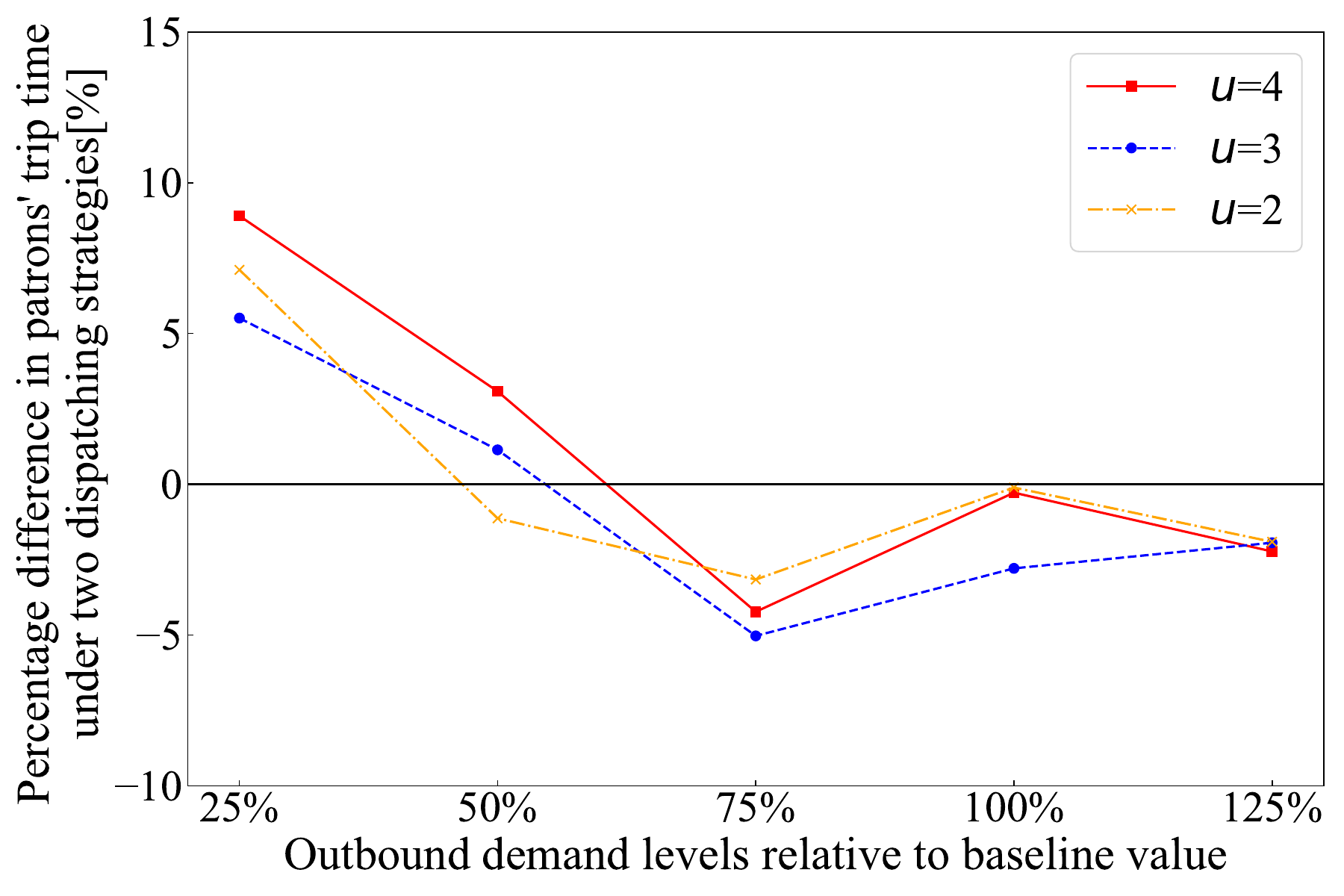}\label{fig:diapatch_uniform}}
        \subfigure[Non-uniform demand scenarios]{\includegraphics[width=0.48\textwidth]{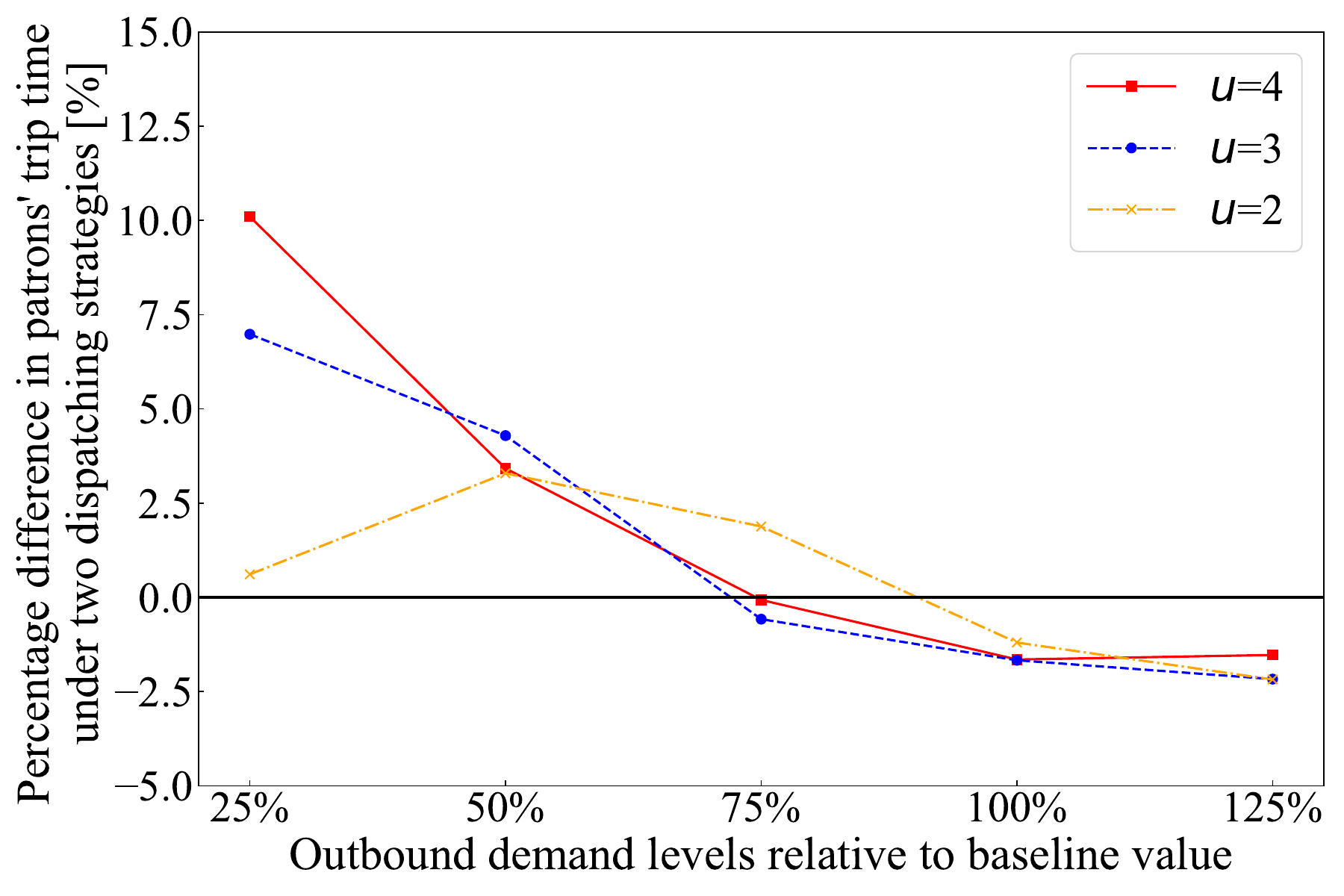}\label{fig:dispatch_non_uniform}}
		\caption{{Percentage difference in patrons' trip time under soft- and hard-target dispatching algorithms.}}\label{fig:soft_target_result}
    \end{figure}

The effect of varying demand levels is examined via sensitivity analyses while keeping all other factors the same as the baseline setting. Figures \ref{fig:soft_target_result}(a) and (b) depict the two dispatching algorithms' percentage differences in patrons' trip time, measured by $ \left(1-\frac{\text{avg}\left(\left\{T^{\rm ST}_i\right\}\right)}{\text{avg}\left(\left\{T^{\rm HT}_i\right\}\right)}\right)\times 100\%$, where avg$(\cdot)$ returns the average of the argument, and the superscripts ST and HT indicate soft- and hard-target algorithms, respectively.

The findings suggest that under low demand levels, such as 25\% of the baseline value, the soft-target algorithm can be more efficient, resulting in considerably shorter trip time for patrons, while the hard-target algorithm excels with medium to high demand levels. The result aligns with \cite{fan2024}'s finding in shuttle transit. {Figure \ref{fig:soft_target_result} also shows the impact of occupancy target $u$. It implies that inappropriate values of $u$ may invert the relative advantage/disadvantage relationship between the two dispatching strategies.}
	
\subsubsection{Repositioning algorithm}\label{sec_repositioning}
This section assesses the urgency concept in the repositioning algorithm. We compare the baseline setting (i.e., $\alpha=0.5$ in Eq. (\ref{urgency})) with two alternative repositioning algorithms: one prioritizing only requests' unmatched time (i.e., $\alpha = 1$) and the other focusing solely on the unmatched requests' distance to the vehicle (i.e., $\alpha = 0$).

{Table \ref{tab:reposition} presents the baseline scenario results for repositioning algorithms with $\alpha=1, 0.5, 0$. While their differences in service rates and patrons' trip time components are small under uniform demand scenarios, they can be notable in non-uniform scenarios. For instance, the service rate of RPaF with $\alpha = 0.5$ exceeds that with $\alpha = 1$ by over 4\%.} The outcome highlights the importance of selecting an appropriate $\alpha$ value in balancing unmatched requests’ waiting times and distance to available vehicles. 
\begin{table}[!ht]
    \centering
    \caption{{Performance of RPaF with varying repositioning algorithms under the baseline scenario.}}
    \label{tab:reposition}
    \begin{tabular}{c|c|c|c|c|c}
         \hline
         \makecell{\textbf{Demand} \\ \textbf{distribution}}  & \textbf{Parameter $\alpha$} & \makecell{\textbf{Service} \\ \textbf{rate [\%]}}
         & \makecell{\textbf{Waiting} \\ \textbf{time [h]}} & \makecell{\textbf{In-vehicle} \\ \textbf{time [h]}} & \makecell{\textbf{Trip} \\ \textbf{time [h]}}  \\
         \hline
         \multirow{3}{*}{Uniform} & 0 & 90 & 0.12 & 0.25 & 0.37 \\
         & 0.5 & 91 & 0.12 & 0.25 & 0.37 \\
         & 1 & 90.5 & 0.13 & 0.25 & 0.38 \\
         \hline
         \multirow{3}{*}{Non-uniform} & 0 & 89.5 & 0.11 & 0.21 & 0.32 \\
         & 0.5 & 93 & 0.12 & 0.21 & 0.33 \\
         & 1 & 88.5 & 0.12 & 0.21 & 0.33 \\
         \hline
    \end{tabular}
    
\end{table}



\subsection{Comparison with two counterpart feeder services} \label{counterparts}
\begin{figure}[!ht]
    \centering
    \includegraphics[width=0.7\textwidth]{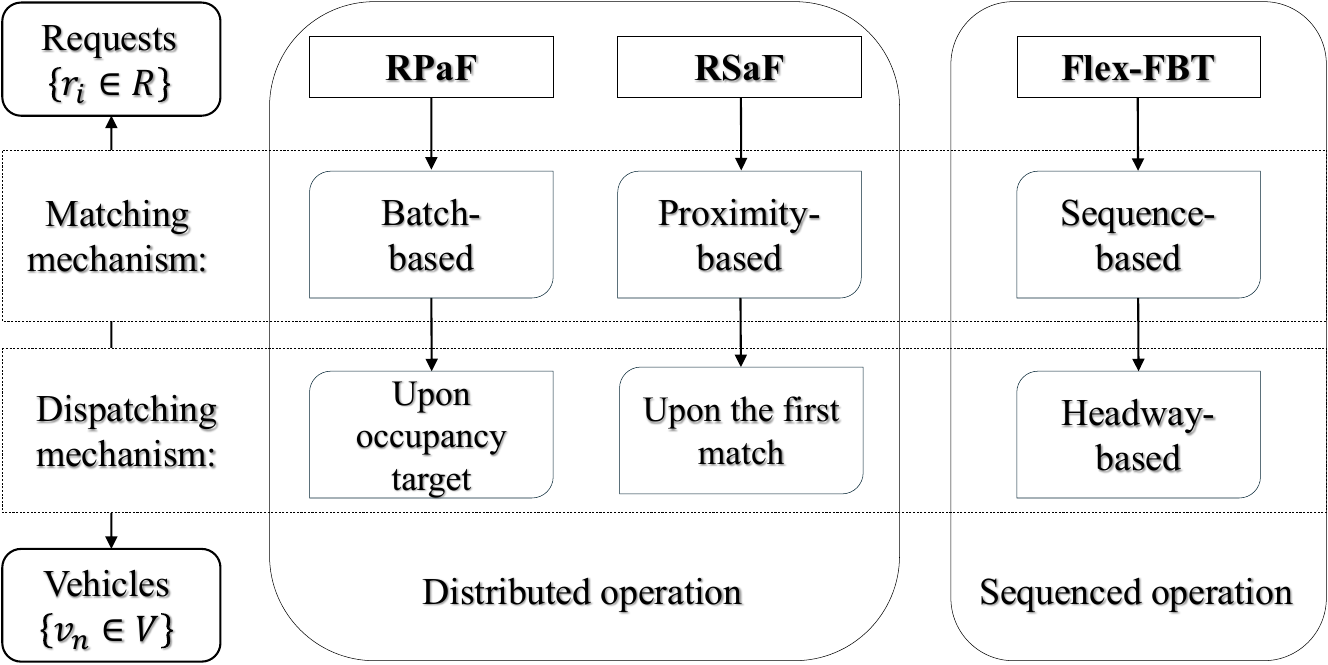}
    \caption{{Schematic diagrams of matching and dispatching mechanisms in RPaF, RSaF, and Flex-FBT services.}}
    \label{fig_schematic_diagram}
\end{figure}

We now start comparing the proposed RPaF with two counterpart feeder services: (\romannumeral 1) the Ride-Sharing as Feeder (RSaF) \citep{DAGANZO2019213, LIU202122}, and (\romannumeral 2) the Flexible-Route Feeder-Bus Transit (Flex-FBT) \citep{chang1991, EDWARDKIM201967}. {The core distinction among them lies in their matching and dispatching mechanisms as illustrated in Figure \ref{fig_schematic_diagram}.}

Nevertheless, the studies on RSaF and Flex-FBT are mostly analytical system designs under idealized assumptions (e.g., spatially uniform demand distribution). For comparability, we devise simulation processes according to their operational characteristics, as given below:
	\begin{enumerate}[label=(\roman*)]
	    \item The RSaF does not pool requests but \textit{instantly} matches and dispatches the available vehicle to pick up the nearest (outbound) request. The vehicle, en route to the initial request, remains available to receive new request assignments until the first passenger boards or the number of assignments reaches the occupancy target $u$. (Thus, we have $o_n \le u, \forall n$.) If multiple requests are matched to the vehicle, it follows a proximity-based routing rule (rather than the TSP plan as in our RPaF), by which it always visits the next closest point. {The inbound operation of RSaF is identical to the proposed RPaF.}
	    \item Flex-FBT dispatches vehicles at regular headways and matches the (outbound) requests within each headway to the successive vehicles. The routing of vehicles follows the optimal TSP-tour plan.
	\end{enumerate}

All parameter values of the two systems, without being explicitly mentioned, are maintained at the baseline setting.  
The simulation steps of RSaF and Flex-FBT are outlined in \ref{appdx-QD} and \ref{appdx-flex}, respectively.
{The following comparisons evaluate the three systems’ performance regarding service rates, patrons' average trip times, and fleet sizes.}

\begin{figure}[!ht]
    \centering
    \subfigure[Uniform demand scenarios]{\includegraphics[width=0.48\textwidth]{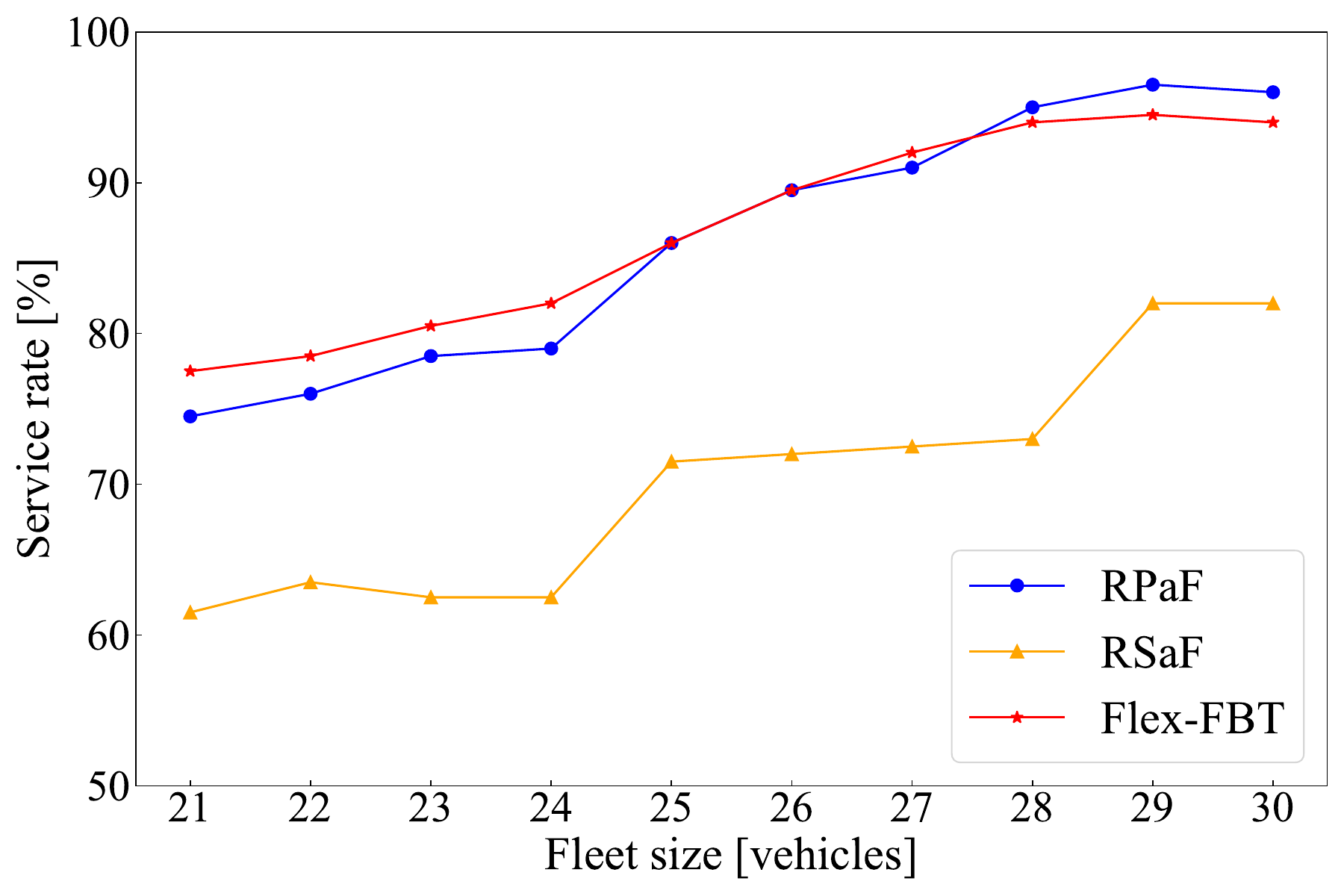}\label{fig:against_service_uniform}}
    \subfigure[Non-uniform demand scenarios]{\includegraphics[width=0.48\textwidth]{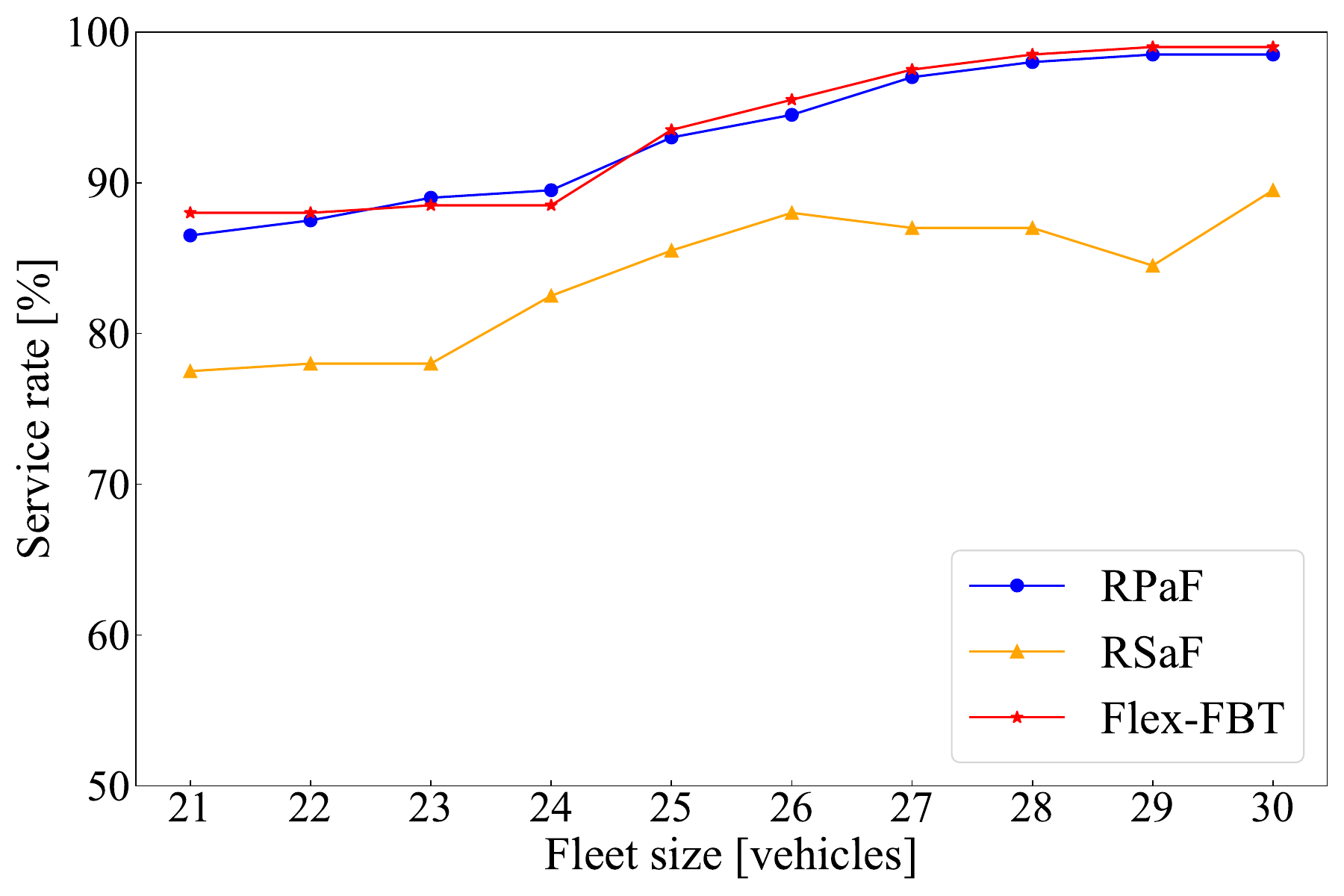}\label{fig:against_service_non_uniform}}
    \caption{{Service rates of RPaF, RSaF, and Flex-FBT under the baseline scenario.}}
    \label{fig:comparison_three_service_rate}
\end{figure}

{Figures \ref{fig:comparison_three_service_rate}(a) and (b) display the service rates of three feeder services under uniform and non-uniform demand scenarios, respectively. As expected, their service rates increase with fleet size. However, for the same fleet size, RPaF and Flex-FBT demonstrate higher service rates than RSaF, due to their greater transportation efficiency.}

{Table \ref{tab:three feeder} summarizes the results for three feeder systems operating at the same service rate, i.e., 90\%. It is not surprising that RSaF requires the largest fleet sizes across most demand scenarios. For patrons in RSaF, however, they enjoy shorter average trip times than those in the other two alternatives. The opposite is Flex-FBT, which usually operates smaller fleet sizes but yields the longest average trip times. The performance of RPaF sits in between. It produces average trip times in the neighborhood of RSaF's while utilizing fewer fleets.}



\begin{table}[!ht]
    \centering
    \caption{{Performance of RPaF, RSaF, Flex-FBT at service rate of 90\%.}}
    \label{tab:three feeder}
    \begin{tabular}{c|c|c|c|c|c|c|c}
         \hline
         \multirow{2}{*}{\makecell{\textbf{Demand}\\ \textbf{distribution}}} & \multirow{2}{*}{\makecell{\textbf{OB demand levels} \\ \textbf{relative to baseline}}}
         & \multicolumn{3}{c|}{\makecell{\textbf{Trip time} \textbf{[h]}}} & \multicolumn{3}{c}{\makecell{\textbf{Fleet size} \textbf{[vehicles]}}} \\
         \cline{3-8}
         & & RPaF & RSaF & Flex-FBT & RPaF & RSaF & Flex-FBT\\
         \hline
         \multirow{5}{*}{Uniform}& 25\% & 0.39 & 0.31 & 0.57 & 15 & 19 & 12\\
         &50\% & 0.40 & 0.29 & 0.55 & 19 & 28 &18 \\
         &75\% & 0.38 & 0.32 & 0.54 & 24 & 30 &24 \\
         &100\% & 0.38 & 0.33 & 0.55 & 27 & 33 &27 \\
         &125\% & 0.38 & 0.35 & 0.54 & 32 & 34 &36 \\
         \hline
         \multirow{5}{*}{Non-uniform}&25\%  & 0.34 & 0.25 & 0.50 & 14 & 18 &10 \\
         &50\% & 0.33 & 0.26 & 0.46 & 16 & 25 & 15 \\
         &75\% & 0.33 & 0.28 & 0.46 & 20 & 27 & 19\\
         &100\% & 0.33 & 0.29 & 0.46 & 25 & 32 & 25\\
         &125\% & 0.33 & 0.30 & 0.45 & 29 & 33 & 33\\
         \hline
    \end{tabular}
    
\end{table}

\section{Real case study}\label{sec:real}
\begin{figure}[!htbp]
		\centering
		\subfigure[Research area (19.9 sq. km)]{\includegraphics[width=0.45\textwidth]{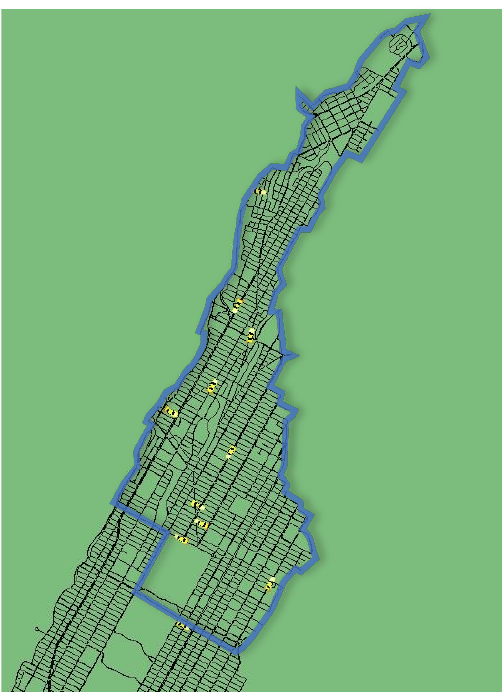}\label{fig:mannet}}
		\hfill
		\subfigure[Outbound demand flow]{\includegraphics[width=0.45\textwidth]{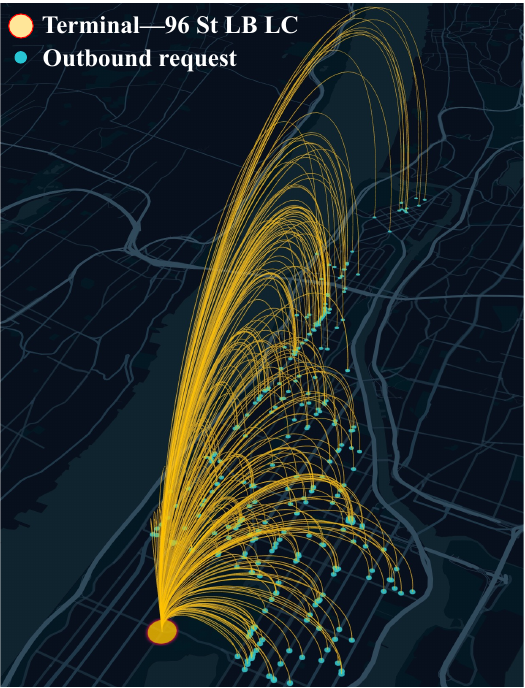}\label{fig:manoutbound}}
		
		\subfigure[Inbound demand flow]{\includegraphics[width=0.45\textwidth]{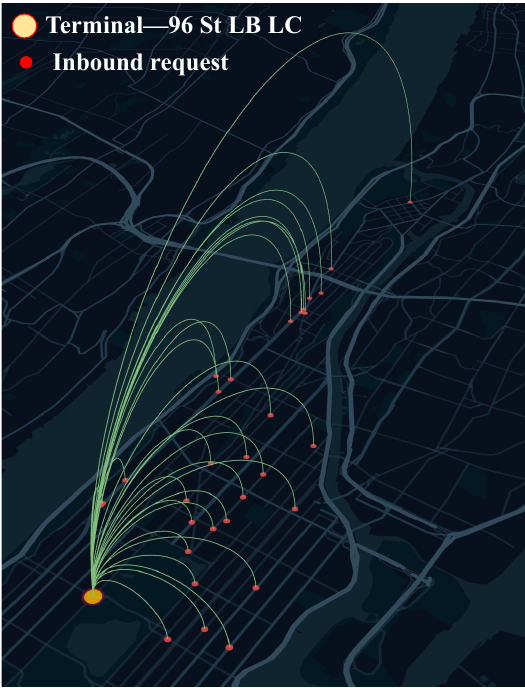}\label{fig:maninbound}}
        \hfill
        \subfigure[Zoning scheme]{\includegraphics[width=0.45\textwidth]{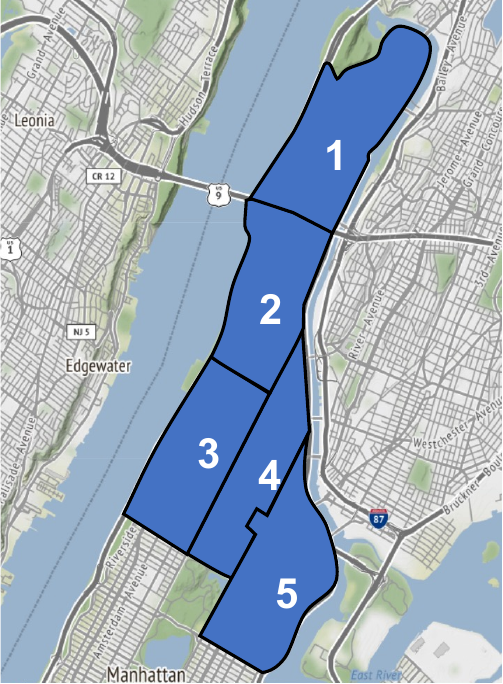}\label{fig:manzone}}
			
		\caption{Visualization of the Manhattan road network and demand data.}
		\label{fig:manhattan}
	\end{figure}

{This section illustrates an application of the proposed RPaF system in a real city, i.e., New York City, USA. The study area occupies the uptown area of Manhattan, roughly 19.9 sq. km, as shown in Figure \ref{fig:mannet}. We retrieve the street network from the OpenStreetMap (\url{https://www.openstreetmap.org}). The network contains 11,503 roadway links and 7,725 intersections. The related network configurations include the length, number of lanes, and average speed of each link, the control types and signal plans (if signalized) of each intersection, one-way streets, turn restrictions, etc. }

{The demand data is sourced from the New York City Taxi Public Dataset \citep{nyc_taxi_dataset}. We extracted the morning-peak trip data (from 7 to 9 a.m.) of green taxis on March 2, 2014. We filter the data with trip ends between uptown Manhattan and a subway transfer station at 96 St LB LC, midtown (where two subway lines intersect). Figures \ref{fig:manoutbound} and Figure \ref{fig:maninbound} visualize outbound and inbound demand, respectively. During the 2-hour study period, there are 321 records of requests (289 for outbound and 32 for inbound). According to the clusters of demand, we partition service zones as shown in Figure \ref{fig:manzone}.}

\begin{figure}
    \centering
    \includegraphics[width=0.65\textwidth]{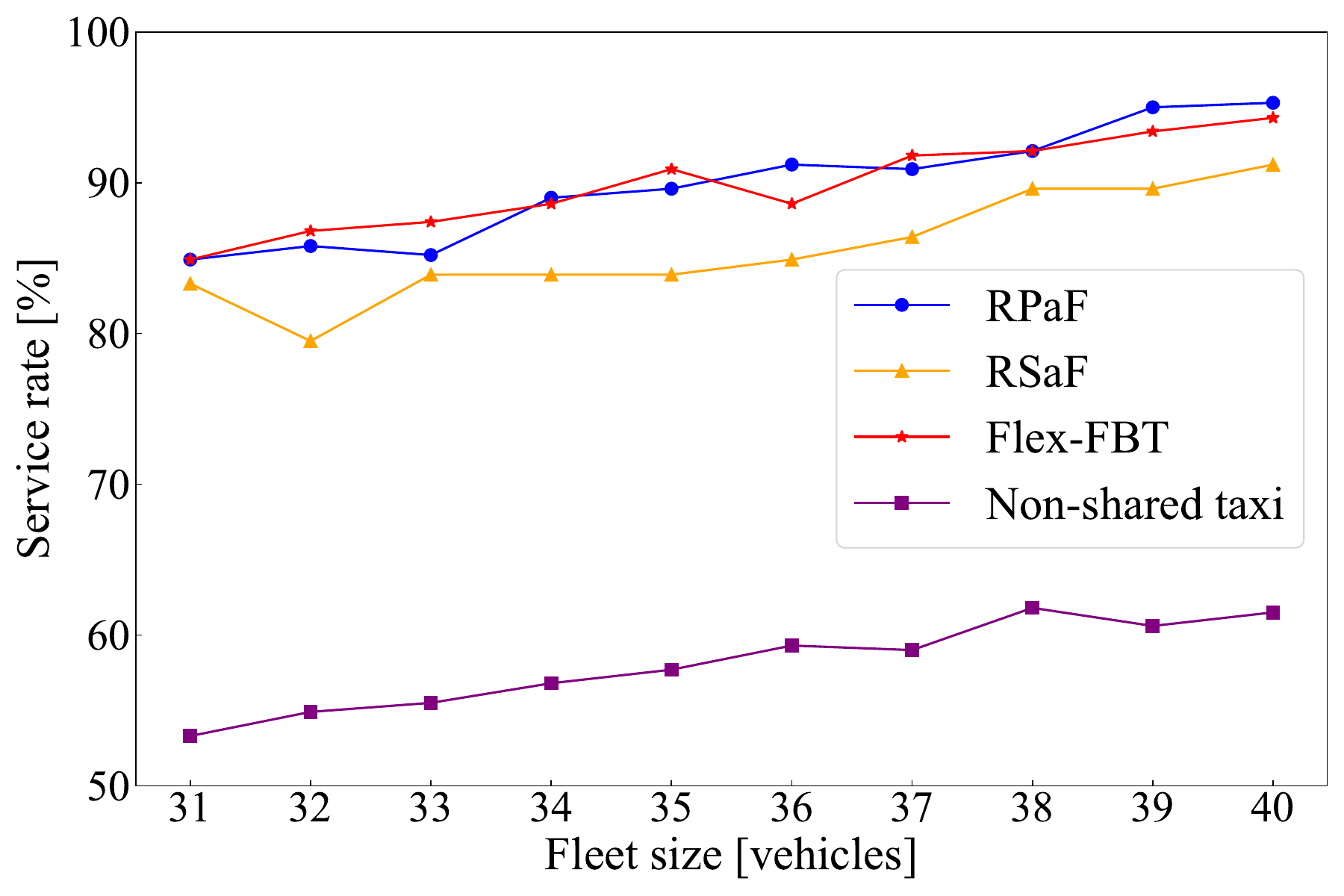}
    \caption{{Service rates of RPaF, RSaF, Flex-FBT, and non-shared taxis with varying fleet sizes.}}
    \label{fig:manhattam_service_rate}
\end{figure}

{We compare three new feeder systems (i.e., RPaF, RSaF, and Flex-FBT) with the existing non-shared taxi service. Figure \ref{fig:manhattam_service_rate} gives the service rates of the four systems with varying fleet sizes. The results imply that all new systems perform better transportation efficiency (with considerably higher service rates per hour) than the existing non-shared taxis. The relationship among RPaF, RSaF, and Flex-FBT aligns with the previous findings.}


{We next scale the current demand data from 25\% to 200\%. For each demand scenario, we adjust the fleet sizes of the three new systems to maintain a service rate of 90\%. The non-shared taxi's fleet size is set to 100 vehicles in all scenarios, ensuring a service rate close to 100\% at current or lower demand levels. This makes our comparison conservative, and the results are summarized in Table \ref{tab:manha}.}

{We have several new findings. First, taxi trips consistently have the shortest average travel time. However, their service rate declines significantly as demand increases, leading to a substantial market gap --- up to 34\% (i.e., $100\%-66\%$) when demand reaches 200\% of current levels. Second, RSaF demonstrates strong competition with taxis at low demand levels, adding only 0.04 hours (2.4 minutes) per patron while operating about 60 fewer vehicles. Third, RPaF shows improved performance at higher demand levels, exhibiting negligible differences in average trip time compared to RSaF, while achieving a reduction in fleet size by eight vehicles. Lastly, Flex-FBT appears less competitive against other systems since it produces significantly longer average trip times for patrons in all scenarios.} 
\begin{table}[!ht]
    \centering
    \caption{{Performance of RPaF, RSaF, Flex-FBT in Manhattan network at service rate of 90\%.}}
    \label{tab:manha}
    \begin{threeparttable}
    \begin{tabular}{c|c|c|c|c|c|c|c}
         \hline
         \multirow{2}{*}{\makecell{\textbf{OB demand levels} \\ \textbf{relative to baseline}}}
         & \multicolumn{4}{c|}{\makecell{\textbf{Trip time}  \textbf{[h]}}} & \multicolumn{3}{c}{\makecell{\textbf{Fleet size} \textbf{[vehicles]}}} \\
         \cline{2-8}
         & Non-shared taxi\tnote{*} & RPaF & RSaF & Flex-FBT  & RPaF & RSaF & Flex-FBT\\
         \hline
          25\% & 0.40 (100\%) & 0.58 & 0.44 & 0.78 & 23 & 30 & 20\\
          50\% & 0.42 (100\%) & 0.59 & 0.46 & 0.78 & 29 & 34 & 28 \\
          75\% & 0.40 (99\%) & 0.59 & 0.53 & 0.85 & 33 & 36 & 32 \\
          100\% & 0.40 (94\%) & 0.60 & 0.55 & 0.83 & 36 & 40 & 35 \\
          125\% & 0.40 (85\%) & 0.60 & 0.56 & 0.86 & 39 & 45 & 41 \\
          ... & ... & ... & ... & ... & ... & ... & ... \\
          200\% & 0.47 (66\%) & 0.61 & 0.60 & 0.87 & 55 & 63 & 59 \\
         \hline
    \end{tabular}

    \begin{tablenotes}
        \footnotesize
        \item[*] The fleet size of non-shared taxis is set to 100 vehicles and the number in parentheses indicates service rates corresponding to this fleet size.
    \end{tablenotes}
    \end{threeparttable}
\end{table}



\section{Conclusion}\label{sec:conclu}
This paper studies Ride-Pooling as Feeder (RPaF) services, whose distinct operation characteristics motivate us to investigate the potential for lowering system costs in a dynamic, door-to-door routing environment. Specifically, we customize three novel algorithms for vehicle-request matching, vehicle dispatching, and repositioning operations. Analytical analysis is utilized to optimally parameterize the algorithms concerning for example the number of pooling requests and matching buffer distance. An open-source microscopic simulation platform is developed for studying RPaF with the proposed algorithms. 

We have conducted extensive experiments in {morning-peak} scenarios to verify the effectiveness of the proposed operation algorithms across varying demand levels and uniform and non-uniform distribution patterns. The performance of RPaF is further compared with two counterparts, namely the Ride-Sharing as Feeder (RSaF) and Flexible-Route Feeder-Bus Transit (Flex-FBT). The results show that RPaF generally performs better {under medium to high demand levels}, {producing higher service rates with the same fleet size or fewer fleet sizes with comparative average trip times for patrons as compared to its counterparts.} {The divergent performance between Flex-FBT and RP/RSaF is notable and may stem from their fundamentally different operational mechanisms, i.e., the so-called sequenced and distributed operation (see descriptions in Section \ref{sec:intro} and illustration in Figure \ref{fig_schematic_diagram}).} Lastly, we demonstrate the application of RPaF in a real-world study of the uptown Manhattan street network (New York City, USA) using actual taxi trip data. Overall, RPaF shows great potential as an effective feeder system, {balancing well the level of service (in terms of service rate and patrons' average trip time) and operational costs (related to fleet size) under the proper conditions.} 

It is worth noting that the current study contains some limitations that warrant future investigation. First, we did not consider riders' {socio-economic heterogeneity, such as varying values of time and diverse preferences, which do exist in reality but require high-quality survey data to unveil. Once available, these factors} can be incorporated more readily into the simulation platform than in the previous analytical models. {To this end, advanced matching and routing algorithms should be developed based on TSP or VRP (Vehicle Routing Problem) with time window constraints \citep{PaoloSanti2014}. In dynamic contexts, solution efficiency is crucial to these algorithms.} {Second, the suggested zoning strategy is contingent upon the availability of parking resources at the hub, which poses practical challenges, as authorities must also address the demands of private vehicle use and cycling. Achieving an efficient and equitable parking allocation requires further investigation into the diverse needs of all stakeholders.} Third, we have set the vehicles in all service zones to be identical. In contrast, a fleet of mixed-size vehicles may benefit the operation during different periods and for service zones with different distances to the hub. {Forth, driver noncompliance issues are also worth exploring, particularly through incentive-based approaches \citep{QIAN2017208,guda2019your}. Fifth, operational issues in non-equilibrium RPaF systems also deserve attention due to constantly increasing/decreasing demand and significant spatial imbalances. In this regard, insights from previous studies on non-shared ride-sourcing mobility \citep[e.g.,][]{ramezani2018dynamic, nourinejad2020ride, ramezani2023dynamic, valadkhani2023dynamic} may guide the development of solutions for dynamic and imbalanced RPaFs.} Lastly, expanding the single-hub system to transit networks fed by RPaFs certainly merits further exploration. It is intriguing to see how much city-wide benefit such integrated systems could bring. 

\section*{Acknowledgments}
The authors appreciate the editor and two anonymous reviewers for their valuable comments that helped improve this paper. The research was partially supported by {the Key Research and Development Support Program of Chengdu Science and Technology Bureau via grant No. 2022-JB00-00006-GX.} We also thank Mr. Fuchao Wang's help with data visualization. 

\newpage
\appendix
\setcounter{table}{0}
\setcounter{figure}{0}
\setcounter{section}{0}
\setcounter{equation}{0}

\section{Lists of abbreviations and notations} \label{appx_notation}
Abbreviations and notations used in this paper are summarized in Table \ref{tab:Abbreviations} and \ref{tab:notation}, respectively.
\begin{table}[!ht]
\centering
\caption{Abbreviations}
\label{tab:Abbreviations}
\begin{tabular}{l l}
\hline
\textbf{Abbreviation} & \textbf{Description}                     \\
\hline
AV                    & Autonomous vehicle                       \\
DAR, DARP             & Dial-a-ride, dial-a-ride problem         \\
DAT                   & Demand adaptive transit                  \\
Fix-FBT               & Fixed-route feeder bus transit           \\
Flex-FBT              & Flexible-route feeder bus transit        \\
M2M, M2O, O2M           & Many to many, many to one, one to many   \\
MAST                  & Mobility allowance shuttle transit       \\
ODSR                  & On-demand shared-ride                    \\
ODT                   & On-demand transit                        \\
P2P-RSaF              & Peer-to-peer ride-sharing as feeder      \\
RP                    & Ride-pooling                             \\
RPaF                  & Ride-pooling as feeder                   \\
RP-HD                 & Ride-pooling with hold-dispatch strategy \\
RSaF                    & Ride-sharing as feeder                            \\
Semiflex-FBT          & Semi-flexible feeder bus transit         \\
SAVaF                & Shared autonomous vehicle as feeder \\
STaF                  & Shared taxi as feeder                    \\
TNC                   & Transportation network company           \\
TSP                   & Traveling salesman problem              \\
\hline
\end{tabular}
\end{table}

    \begin{table}[!ht]
        \centering
        \caption{List of notations}
        \label{tab:notation}
        \begin{tabular}{p{3.5cm} l l }
             \hline
             \textbf{Variables, sets, parameters} & \textbf{Description} & \textbf{Unit} \\
             \hline
             \multicolumn{3}{l}{{Variables}}\\
             \hline
             $U$ & Pooling target for the number of requests & patrons\\
             $\Delta$ & Matching buffer distance & km\\
             \hline
             \multicolumn{3}{l}{{Patron's set $R \equiv \{r_i=\langle t_i,\Vec{l}_i,\delta_i\rangle\}$}}\\
             \hline
             $\tau_i$ & Patron $i$'s request call-in timestamp & /\\
             $\Vec{l}_i$ & Patron $i$'s location & /\\
             $\delta_i$ & Patron $i$'s indicator variable & /\\
             \hline
             \multicolumn{3}{l}{{Unmatched requests set $\widetilde{X}_0$}}\\
             \hline
             \multicolumn{3}{l}{{Vehicle's set $V \equiv \{v_n=\langle o_n,\Vec{l}_n,X_n,\delta_n \rangle\}$}}\\
             \hline
             $o_n$ & Outbound vehicle $n$'s occupancy & patrons/vehicle\\
             $\tilde{o}_n$ & Inbound vehicle $n$'s occupancy & patrons/vehicle\\
             $\Vec{l}_n$ & Vehicle $n$'s location & /\\
             $X_n$ & The set of requests assigned to the vehicle $n$ & /\\
             $\delta_n$ & Vehicle $n$'s indicator variable & /\\
             \hline
             \multicolumn{3}{l}{{Parameters}}\\
             
             
             \hline
             $c_f$ & Operational cost per dispatch & \$/dispatch\\
             $C$ & Vehicle capacity & seats/vehicle\\
             $k$ & Constant parameter in TSP-tour length formula & /\\
             $u$ & Vehicle occupancy target & patrons/vehicle\\
             $L$ & Line haul distance & km\\
             $N$ & Fleet size & vehicles\\
             $S,S^{'}$ & Vehicle speed on the freeway and suburban streets & km/h \\
             $S^{'}(u)$ & Vehicle commercial speed in TSP tours & km/h \\
             $t_{d}$ & Vehicle delay & seconds\\
             $T_i,T_{i}^{w},T_{i}^{r}$ & Average trip time components of patron $i$ & seconds\\
             $\lambda_{\text{OB}},\lambda_{\text{IB}}$ & Demand density of outbound and inbound patrons & trips/$\text{km}^2$/h\\
             $\beta$ & Patrons' value of time (VOT) & \$/hour/patron \\
             $\eta$ & Parameter in inbound vehicles' trip distance formula & / \\
             $\Bar{\tau}$ & Patrons' maximum tolerance time & minutes \\
             \hline
        \end{tabular}
    \end{table}

\newpage
\section{Proof to Eq. (\ref{approx_trip_time})}\label{appdx_u-U}
Let $D$ be the distance of the random request to the center (i.e., the location of the vehicle) of the area of $A(\Delta)$. The $D$ is i.i.d. and uniformly distributed in the range of $[0, \Delta]$. The $U$ points of $D$ divide the range into $U+1$ segments, and the distance between the $u$-th closest point to the center will be $\frac{u}{U+1}\Delta$, which defines an area of $\left(\frac{u}{U+1}\right)^2A(\Delta)$. In this area, the expected TSP-tour distance of visiting $u$ points is
\begin{align}
    k\sqrt{A(\Delta)u}\frac{u}{U+1},
\end{align}
which yields Eq. (\ref{approx_trip_time}) by dividing the commercial speed $S'(u)$.

\section{{Pseudo-code of RSaF}}\label{appdx-QD}
For comparability, we set RSaF vehicles' occupancy target as $u = 4$ identical to our RPaF's. Note that in the original theoretical studies \citep{DAGANZO2019213, LIU202122}, RSaF was restricted to cases of $u = 2$ for the sake of traceability in solving multiple equations of states, whose number grows quadratically with $u$. Yet, expanding our simulation platform to accommodate higher $u$ values is much simpler. The following provides the pseudo-code for simulating RSaF.
	\begin{breakablealgorithm}
		\caption{Simulation of RSaF}
		\hspace*{0.02in} {\bf Input:} 
		Requests set $ R \equiv \{r_i=\langle t_i,\Vec{l}_i,\delta_i\rangle\} $, vehicles set $ V \equiv \{v_n=\langle o_n,\Vec{l}_n,X_n,\delta_n \rangle\} $, Simulation maximum time $t_{\text{max}}$, simulation time step $\Delta t$.\\
		\begin{algorithmic}[1]
			\State Set simulation time step $t=0$.
			
			\While{$t \leq t_{\text{max}}$}
            \State \textbf{Initialization:}
			\If{$t=0$}
			     \State Any vehicle $n$ goes to the initial located position, awaiting assignment.
			\EndIf
            \State \textbf{Matching:}
			\If{$t$ reach $\Delta t$}
			     \State Apply the proximity-based principle to output vehicle-demand mapping dictionary $\mathbf{X} = \{X_n\}$.
			\EndIf
            \For{each zone}
			\State \textbf{Iterate over vehicles and send commands:}
			\For{each vehicle $n$ in vehicle set $V$} 
            \State \textbf{Outbound service:}
			\If{$v_n=\langle o_n=0,\Vec{l}_n,X_n=\emptyset,\text{OB} \rangle$ state} 
                \State Vehicle $n$ remains in its initial state and stays stationary.
			\ElsIf{$v_n=\langle o_n=0,\Vec{l}_n,X_n\neq \emptyset,\text{OB} \rangle$ state}    
                \State Vehicle $ n $ departs to pick up the first patron.
            \ElsIf{$v_n=\langle o_n=1,\Vec{l}_n,X_n,\text{OB} \rangle$ state}
                \State The first patron is already on board.
                \State Vehicle $n$ stops matching.
                \If{$X_n=\emptyset$}
                    \State Vehicle $n$ goes directly to the hub.
                \ElsIf{$X_n \neq \emptyset$}
                    \State Vehicle $n$ picks up other matched requests $X_n$, and finally returns to the hub.
                \EndIf
            \ElsIf{vehicle $n$ arrives at the hub}
                \State Outbound patrons alight.
                \State \textbf{Inbound service:}
                \State $v_n \leftarrow v_n=\langle \tilde{o}_n,\Vec{l}_n,X_n,\delta_n=\text{IB} \rangle$.
                \If{$\tilde{o}_n=0$}
                    \State Vehicle $n$ returns the last patron pick-up position $\Vec{l}^{'}$ in the outbound service.
                \ElsIf{$\tilde{o}_n>0$}
                    \State Vehicle $n$ sends inbound patrons back to the suburbs under the routing algorithm until the last drop-off position $\Vec{l}^{'}$.
                \EndIf
            \ElsIf{$v_n=\langle \tilde{o}_n=0,\Vec{l}_n=\Vec{l}^{'},X_n=\emptyset
					,\text{IB} \rangle$}
                \State $v_n \leftarrow v_n=\langle o_n,\Vec{l}_n,X_n,\delta_n=\text{OB} \rangle$.
			\EndIf
			
			\EndFor
            \EndFor
			\If{all patrons have been served}
			\State \textbf{break}
			\EndIf
			\State $t=t+\Delta t$.
			\EndWhile
			
		\end{algorithmic}
	\end{breakablealgorithm}

\section{Pseudo-code of Flex-FBT}\label{appdx-flex}
The headway of Flex-FBT, denoted by $h$, is optimized using the models in Appendix D of \cite{fan2024optimal}, set to 9.42 minutes under the baseline setting and updated for every other scenario. The detailed optimization models are omitted for brevity. The pseudo-code for simulating Flex-FBT is provided below.
	\begin{breakablealgorithm}
    	\caption{Simulation of Flex-FBT}
    	\label{algorithm4}
    	\hspace*{0.02in} {\bf Input:} 
    	Requests set $ R \equiv \{r_i=\langle t_i,\Vec{l}_i,\delta_i\rangle\} $, vehicles set $V \equiv \{v_n=\langle o_n,\Vec{l}_n,X_n,\delta_n \rangle\}$, simulation maximum time $t_{\text{max}}$, fleet size $N$, vehicle capacity $C$, optimal headway $h$, simulation time step $\Delta t$, number of sub-areas $A$.\\
    	\begin{algorithmic}[1]
            \State \textbf{Initialization:}
            \State Set simulation time step $t=0$.
            \State Set initial departure time $t_p=h$.
    		\State Define $R_{\text{OB}}=\emptyset$ as the set of outbound patrons to be matched in the suburb.
    		\State Define $R_{\text{IB}}=\emptyset$ as the set of inbound patrons waiting at the hub.
            \While{$t \leq t_{\text{max}}$}
                \For{each zone id $a \in \{1,...,A\}$}
    		    \State Renew $V^{(a)}=\{v_n\}$, $R^{(a)}_{\text{OB}}$, $R^{(a)}_{\text{IB}}$.
    		    \If{$|V^{(a)}|<N^{(a)}$, where $|\cdot|$ represents the number of service vehicles in zone $a$}
                    \If{$R^{(a)}_{\text{IB}}=\emptyset$}
                        \State \textbf{No inbound requests:}
                        \If{$|R^{(a)}_{\text{OB}}| \geq C$}
                            \State \textbf{Reach full match:}
                            \State Filter the first $\text{min}\{ C,|R^{(a)}_{\text{OB}}| \}$ elements in $R^{(a)}_{\text{OB}}$ and determine them be set $X_{n}$.
                            \State Dispatch vehicle $n$ with state $v_n=\langle \tilde{o}_n=0,\Vec{l}_n,X_n,\delta_n=\text{OB} \rangle$.
                            \State Vehicle $n$ picks up matched outbound requests under the instructions of the routing algorithm and finally returns to the hub.
                            \State $V^{(a)} \leftarrow V^{(a)} \cup v_n$.
                        \ElsIf{$t$ reach $t^{(a)}_p$ and $R^{(a)}_{\text{OB}} \neq \emptyset$}
                            \State \textbf{Reach departure interval:}
                            \State $X_n=R^{(a)}_{\text{OB}}$.
                            \State Dispatch vehicle $n$ with state $v_n=\langle \tilde{o}_n=0,\Vec{l}_n,X_n,\delta_n=\text{OB} \rangle$.
                            \State Vehicle $n$ picks up matched outbound requests under the instructions of the routing algorithm and finally returns to the hub.
                            \State $V^{(a)} \leftarrow V^{(a)} \cup v_n$.
                        \EndIf
                    \State $t^{(a)}_{p}=t+h$.
                    \ElsIf{$R^{(a)}_{\text{IB}} \neq \emptyset$}
                        \State \textbf{Have inbound requests:}
                        \State Filter the first $\text{min}\{ C,|R^{(a)}_{\text{IB}}| \}$ elements in $R^{(a)}_{\text{IB}}$ and let them be Set $X_{\rm IB}$.
                        \If{$|R^{(a)}_{\text{OB}}| \geq C$}
                            \State \textbf{Reach full match:}
                            \State Filter the first $\text{min}\{ C,|R^{(a)}_{\text{OB}}| \}$ elements in $R^{(a)}_{\text{OB}}$ and determine them be set $X_{\rm OB}$.
                            \State Dispatch vehicle $n$ with state $v_n=\langle \tilde{o}_n>0,\Vec{l}_n,X_n=X_{\rm IB} \cup X_{\rm OB},\delta_n \rangle$.
                            \State Generate inbound route $\Upsilon_{\text{IB}}$ based on $X_{\rm IB}$; and generate outbound route $\Upsilon_{\text{OB}}$ based on $X_{\rm OB}$.
                            \State Vehicle $n$ ($\delta_n=\text{IB}$) first delivers inbound patrons along the inbound route $\Upsilon_{\text{IB}}$; then switches state ($\delta_n=\text{OB}$) to pick up outbound patrons along the outbound route $\Upsilon_{\text{OB}}$; and finally returns to the hub.
                            \State $V^{(a)} \leftarrow V^{(a)} \cup v_n$.
                        \ElsIf{$t$ reach $t^{(a)}_p$ and $R^{(a)}_{\text{OB}} \neq \emptyset$}
                            \State \textbf{Reach departure interval:}
                            \State $X_{\rm OB}=R^{(a)}_{\text{OB}}$
                            \State Dispatch vehicle $n$ with state $v_n=\langle \tilde{o}_n>0,\Vec{l}_n,X_n=X_{\rm IB}\cup X_{\rm OB},\delta_n=\text{IB} \rangle$.
                            \State Generate inbound route $\Upsilon_{\text{IB}}$ based on $X_{\rm IB}$; and generate outbound route $\Upsilon_{\text{OB}}$ based on $X_{\rm OB}$.
                            \State Vehicle $n$ ($\delta_n=\text{IB}$) first delivers inbound patrons along the inbound route $\Upsilon_{\text{IB}}$; then switches state ($\delta_n=\text{OB}$) to pick up outbound patrons along the outbound route $\Upsilon_{\text{OB}}$; and finally returns to the hub.
                            \State $V^{(a)} \leftarrow V^{(a)} \cup v_n$.
                            
                        \EndIf
                    \State $t^{(a)}_{p}=t+h$.
                    \EndIf
                \EndIf
                \For{each vehicle $n$ in vehicle set $V^{(a)}$}
                    \If{vehicle $n$ returns to hub}
                        \State Renew $V^{(a)} \leftarrow V^{(a)}  \setminus \{v_n\}$.
                    \EndIf
                \EndFor
                \EndFor
        		\If{all patrons have been served}
        		    \State \textbf{break}.
        		\EndIf
    		    \State $t=t+\Delta t$.
    		\EndWhile
    	\end{algorithmic}
    \end{breakablealgorithm}

\section{Open-tour TSP model} \label{open-tour tsp}
We employ the classic open-tour TSP model (\ref{eq.route}) to generate the optimal route for visiting $|X_n|$ request points for any vehicle $n$ \citep{sengupta2019, singamsetty2021}. We acquire the match set $X_n$ as input to the model. The current location of vehicle $n$ is represented by $\Vec{l}_n$, and the locations of all matched requests are {$\{\Vec{l}_{i \in X_n}\}$}. They together constitute the set of to-be-visit points, denoted $\Theta$ (where subscript $n$ is dropped without confusion). The total number of points in the set is thus $\theta=|\Theta| \le |X_n|+1$, which takes the equality sign when $\Vec{l}_n$ does not overlap with any of $\{\Vec{l}_{i \in X_n}\}$. Additionally, we retrieve the travel distances or costs between the points, denoted $\{c_{ij}: i, j \in \Theta\}$, from the simulation environment. 

We introduce: binary decision variables $\{x_{ij} \in \{0,1\}: i, j \in \Theta\}$, where $x_{ij}=1$ means the edge $(i,j)$ is included in the tour, and otherwise $x_{ij}=0$; and auxiliary decision variables $\{d_{k}: k \in \Theta \}$, where $d_{k}=-1$ denotes the start point, $d_{k}=+1$ the endpoint, and $d_{k}=0$ intermediate points. Since the start point is known, we can stipulate the rest $d_{k}$ as binary variables and enforce that their sum is equal to 1 \citep{rasmussen2011tsp}.
Therefore, the open-tour TSP can be constructed to minimize the total distance or cost, as given by
    \begin{subequations}\label{eq.route}
        \begin{align}
            \minimize_{\{x_{ij}\}, \{d_{k}\}} & \quad \sum_{i=1}^{\theta}\sum_{j=1}^{\theta} c_{ij} x_{ij}\label{route:obj},
        \end{align}
        subject to:
        \begin{align}
            & \quad \sum_{i=1}^{\theta} x_{ij} = 1, \quad \forall j>1, j\in \Theta \label{route:cons1}, \\
            & \quad \sum_{i=1}^{\theta} x_{ik} - \sum_{j=1}^{\theta} x_{kj}=d_{k}, \quad \forall k \in \Theta \label{route:cons2},\\
            & \quad d_{1}=-1, \sum_{k=2}^{\theta} d_{k}=1, \label{route:cons3}\\
            & \quad \sum_{i\in M}\sum_{j\in M}x_{ij}\leq |M|-1,\quad \forall M\subset \Theta, 2 \leq |M| \leq \theta-2, \label{route:cons4}\\
            & \quad x_{ij} \in \{0,1\},\quad \forall i,j\in \Theta, \label{route:cons5}\\
            & \quad d_{k} \in \{0,1\},\quad \forall k>1, k \in \Theta, \label{route:cons6}
        \end{align}
    \end{subequations}
where constraints (\ref{route:cons1}) ensure each point is visited once; constraints (\ref{route:cons2}) guarantee that the vehicle leaves from the start point, visits all intermediate points, and finishes at the endpoint; constraint (\ref{route:cons3}) ensures the open tour, meaning the vehicle does not return to the start point; constraint (\ref{route:cons4}) avoids subtours; and constraints (\ref{route:cons5}) and (\ref{route:cons6}) define the valid range of decision variables.



\section{{Operational algorithm for general trips completed within the suburb}} \label{app_RM}
\begin{figure}[!ht]
    \centering
    \includegraphics[width=0.5\linewidth]{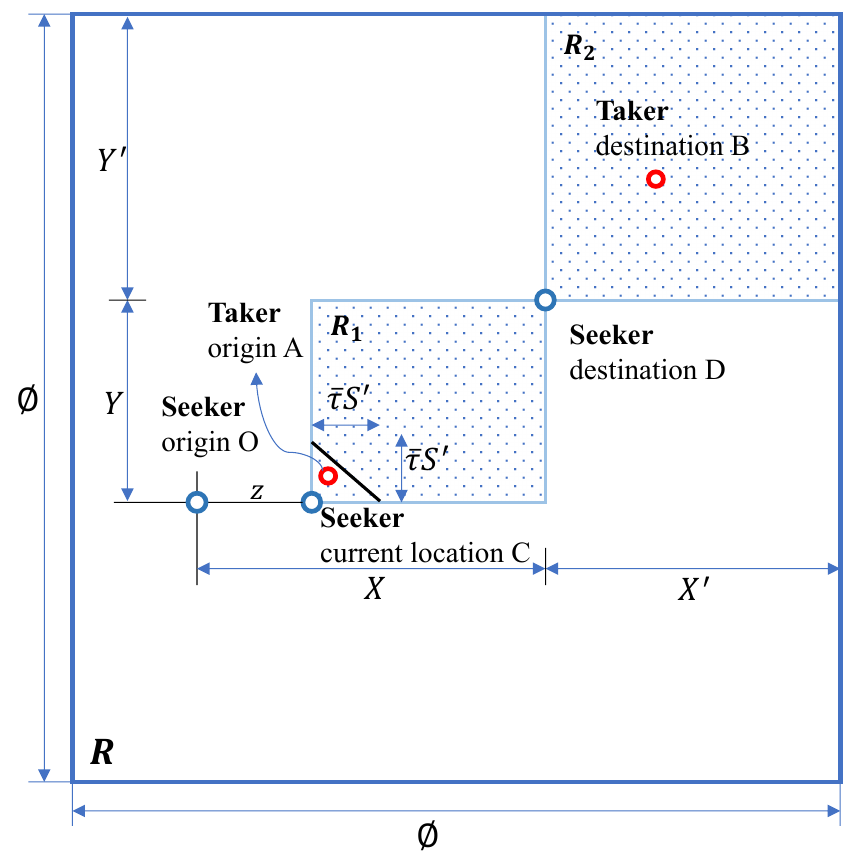}
    \caption{Illustration of Ride-Matching for general trips (\cite{daganzo2019public})}
    \label{fig:general_RM}
\end{figure}

{For illustration, Figure \ref{fig:general_RM} depicts a Ride-Matching (RM) scenario between two riders, namely Seeker and Taker. Without loss of generality, the vehicle is assumed to travel horizontally first, followed by a vertical movement. Within the service zone $R$, assume that Seeker boarded the vehicle at point O and traveled $z$ km to reach the current location C. The sub-region formed by points C and D is defined as $R_{1}$, while the sub-region formed by D and the boundary of $R$ is defined as $R_{2}$. To identify a valid RM, we verify Taker's Origin-Feasibility (OF) and Destination-Feasibility (DF) based on two criteria: (i) Taker can be picked up within her maximum tolerance time $\Bar{\tau}$, and (ii) for combined trips, neither rider requires a detour.} 

{The steps of RM operation are detailed in Algorithm \ref{algorithm: general RM}:
}

\begin{breakablealgorithm}
    \caption{{Ride-Matching algorithm for general trips}}
    \label{algorithm: general RM}
    \makebox[\textwidth][l]{\bf Input: {Vehicle speed $S'$, matching tolerance time $\Bar{\tau}$}.}
    \begin{algorithmic}[1]
        \State Attain OD information (O, D) and current location C of Seeker.
        \While{the horizontal distance $z$ traveled by the vehicle is less than the horizontal distance between D and O, denoted $X$}
            \State New request arises (Taker).
            \State The origin and destination of the Taker are A and B, respectively.
            \State \textbf{Initialization:}
            \State OF, DF = \textbf{False}, \textbf{False}
            \State \textbf{Check Origin-Feasibility (OF):}
            \If{the Manhattan distance between A and C is less than or equal to $\Bar{\tau} S'$ (i.e. A is located in the isosceles right triangle area)}
                \State OF=\textbf{True}
            \EndIf
            \State \textbf{Check Destination-Feasibility (DF):}
            \If{B is located in the upper right of A in the region $R_{1}$ or in the region $R_{2}$}
                \State DF=\textbf{True}
            \EndIf
            \State \textbf{Check valid ride-matching:}
            \If{OF \&\& DF == \textbf{True}}
                \State Seeker and Taker are successfully matched.
            \EndIf
        \EndWhile
    \end{algorithmic}
\end{breakablealgorithm}

\newpage
\bibliographystyle{elsarticle-harv} 
\bibliography{refer}






\end{document}